\documentstyle[12pt,psfig,epsfig,axodraw]{article}
\advance\textheight by \topskip
\newcommand{\newc}{\newcommand}

\newcommand{\be}{\begin{equation}}
\newcommand{\ee}{\end{equation}}
\newcommand{\br}{\begin{eqnarray}}
\newcommand{\er}{\end{eqnarray}}
\newcommand{\ba}{\begin{array}}
\newcommand{\ea}{\end{array}}
\newcommand{\bi}{\begin{itemize}}
\newcommand{\ei}{\end{itemize}}
\newcommand{\bn}{\begin{enumerate}}
\newcommand{\en}{\end{enumerate}}
\newcommand{\bc}{\begin{center}}
\newcommand{\ec}{\end{center}}

\newcommand{\ar}{\rightarrow}

\newcommand{\Dir}{\kern -6.4pt\Big{/}}
\newcommand{\Dirin}{\kern -10.4pt\Big{/}\kern 4.4pt}
\newcommand{\DDir}{\kern -10.6pt\Big{/}}
\newcommand{\DGir}{\kern -6.0pt\Big{/}}
\def\Ecm{\ifmmode{E_{\mathrm{cm}}}\else{$E_{\mathrm{cm}}$}\fi}
\def\gluino{\ifmmode{\mathaccent"7E g}\else{$\mathaccent"7E g$}\fi}
\def\photino{\ifmmode{\mathaccent"7E \gamma}\else{$\mathaccent"7E \gamma$}\fi}
\def\mgluino{\ifmmode{m_{\mathaccent"7E g}}
             \else{$m_{\mathaccent"7E g}$}\fi}
\def\taugluino{\ifmmode{\tau_{\mathaccent"7E g}}
             \else{$\tau_{\mathaccent"7E g}$}\fi}
\def\mphotino{\ifmmode{m_{\mathaccent"7E \gamma}}
             \else{$m_{\mathaccent"7E \gamma}$}\fi}
\def\ML{\ifmmode{{\mathaccent"7E M}_L}
             \else{${\mathaccent"7E M}_L$}\fi}
\def\MR{\ifmmode{{\mathaccent"7E M}_R}
             \else{${\mathaccent"7E M}_R$}\fi}
\def\lsim{\buildrel{\scriptscriptstyle <}\over{\scriptscriptstyle\sim}}
\def\gsim{\buildrel{\scriptscriptstyle >}\over{\scriptscriptstyle\sim}}
\def\issue(#1,#2,#3){{\bf #1}, #2 (#3)} 
\def\PRD(#1,#2,#3){Phys.\ Rev.\ D \issue(#1,#2,#3)}
\def\NPB(#1,#2,#3){Nucl.\ Phys.\ B \issue(#1,#2,#3)}
\def\JP(#1,#2,#3){J.\ Phys.\issue(#1,#2,#3)}
\def\PL(#1,#2,#3){Phys.\ Lett. \issue(#1,#2,#3)}
\def\PLB(#1,#2,#3){Phys.\ Lett.\ B  \issue(#1,#2,#3)}
\def\ZP(#1,#2,#3){Z.\ Phys. \issue(#1,#2,#3)}
\def\ZPC(#1,#2,#3){Z.\ Phys. \ C  \issue(#1,#2,#3)}
\def\PREP(#1,#2,#3){Phys.\ Rep. \issue(#1,#2,#3)}
\def\PRL(#1,#2,#3){Phys.\ Rev.\ Lett. \issue(#1,#2,#3)}
\def\MPL(#1,#2,#3){Mod.\ Phys.\ Lett. \issue(#1,#2,#3)}
\def\RMP(#1,#2,#3){Rev.\ Mod.\ Phys. \issue(#1,#2,#3)}
\def\SJNP(#1,#2,#3){Sov.\ J. \ Nucl.\ Phys. \issue(#1,#2,#3)}
\def\CPC(#1,#2,#3){Comp.\ Phys. \ Comm. \issue(#1,#2,#3)}
\def\IJMPA(#1,#2,#3){Int.\ J. \ Mod. \ Phys.\ A \issue(#1,#2,#3)}
\def\MPLA(#1,#2,#3){Mod.\ Phys.\ Lett.\ A \issue(#1,#2,#3)}
\def\PTP(#1,#2,#3){Prog.\ Theor.\ Phys. \issue(#1,#2,#3)}
\def\RMP(#1,#2,#3){Rev.\ Mod.\ Phys. \issue(#1,#2,#3)}
\def\NIMA(#1,#2,#3){Nucl.\ Instrum.\ Methods \ A \issue(#1,#2,#3)}
\def\JHEP(#1,#2,#3){J.\ High \ Energy \ Phys. \issue(#1,#2,#3)}
\def\EPJC(#1,#2,#3){Eur.\ Phys.\ J. \ C \issue(#1,#2,#3)}
\newc{\PRDR}[3]{{Phys. Rev. D} {\bf #1}, Rapid 
Communications, #2 (#3)}

\def\Ord{\lower .7ex\hbox{$\;\stackrel{\textstyle <}{\sim}\;$}}
\def\OOrd{\lower .7ex\hbox{$\;\stackrel{\textstyle >}{\sim}\;$}}
\def \ch1plus {{\tilde\chi_1^+}}
\def \ch1minus {{\tilde\chi_1^-}}
\def \chplus {{\tilde\chi^+}}
\def \chminus {{\tilde\chi^-}}
\def \ch1pm{{\tilde\chi_1}^{\pm}}

\def \LSP{\tilde\chi_1^0}
\def \LSPsecond{\tilde\chi_2^0}
\def \STAU{\tilde\tau_1}
\def \MSTAU{m_{\tilde\tau_1}}

\def \MSL1{m_{\tilde\l_1}}
\def \MSMU1{m_{\tilde\mu_1}}
\def \SNU{\tilde{\nu}}

\def \MLSP{m_{{\tilde\chi_1}^0}}

\def \MCH{m_{{\tilde\chi}^{\pm}}}

\def \MSNU{m_{\tilde{\nu}}}

\def \MCH{m_{{\tilde\chi}_1^{\pm}}}
\def \LSP{\tilde\chi_1^0}

\def \MSNU{m_{\tilde{\nu}}}

\def \mhf{m_{1/2}}
\def \MST{m_{\tilde t_1}}

\def\t1{\tilde{t_1}}
 
\newc{\rpv}{{\not\!\!R_p}}
\newc{\rpvm}{{\not R_p}}
\newc{\ttbar}{\mbox{$t\overline{t}$}}
\newc{\qbar}     {\mbox{$\overline{q}$}}
\newc{\squark}   {\mbox{$\tilde{q}$}}
\newc{\sqbar}    {\mbox{$\bar{\tilde{q}}$}}
\newc{\msquark}  {\mbox{$M(\squark)$}}
\newc{\csquarkl} {\mbox{$\tilde{c}_L$}}
\newc{\mcsl}     {\mbox{$M(\csquarkl)$}}
\newc{\ssb}      {\mbox{$\squark\overline{\squark}$}}
\newc{\csquark}  {\mbox{$\tilde{c}$}}
\newcommand{\tsquark}  {\mbox{$\tilde{t}$}}

\newc{\ttbone}   {\mbox{$\tsquark_1\overline{\tsquark}_1$}}
\newc{\chione}   {\mbox{$\tilde{\chi}_{1}^{\pm}$}}
\newc{\mchione}  {\mbox{$M(\tilde{\chi}_{1}^{\pm})$}}
\newc{\mstopo}   {\mbox{$M(\tilde{t}_1)$}}
\newc{\mcone}    {\mbox{$M(\tilde{\chi}_{1}^{\pm})$}}
\newc{\none}     {\mbox{$\tilde{\chi}_{1}^0$}}
\newc{\mchio}    {\mbox{$M(\none)$}}
\newc{\lsp}      {\mbox{$\tilde{\chi}_{1}^0$}}
\newc{\mz}       {\mbox{$M_0$}}
\newc{\mo}       {\mbox{$M_{1/2}$}}
\newc{\lamp}     {\mbox{$\lambda_{121}'$}}
\newc{\tev}  {\mbox{$\;{\rm TeV}$}}
\newc{\gevc} {\mbox{$\;{\rm GeV}/c$}}
\newc{\gevcc}{\mbox{$\;{\rm GeV}/c^2$}}
\newc{\chis}{\mbox{$\chi^{2}$}}
\newc{\ifb}{\mbox{${\rm fb}^{-1}$}}
\newc{\ipb}{\mbox{${\rm pb}^{-1}$}}
\newc{\met}{\mbox{${E\!\!\!\!/_T}$}}
\newc{\intlum}{\mbox{${ \int {\cal L} \; dt}$}}
\newc{\et}{\mbox{$E_T$}}
\newc{\modulus}[1]{\left| #1 \right|}
\newc{\bp}{\mbox{$b'$}}
\newc{\lxy}{\mbox{$L_{xy}$}}
\newc{\dedx}{\mbox{${\rm d}E/{\rm d}x$}}
\newc{\R}{$R$}
\newc{\charginom}{M_{\tilde \chi}^{+}}
\newc{\mue}{\mu_{\tilde{e}_{iL}}}
\newc{\mud}{\mu_{\tilde{d}_{jL}}}

\newc{\barr}{\begin{eqnarray}}
\newc{\earr}{\end{eqnarray}}
\newc{\beq}{\begin{equation}}
\newc{\eeq}{\end{equation}}
\newc{\ra}{\rightarrow}
\newc{\lra}{\longrightarrow}
\newc{\lam}{\lambda}
\newc{\eps}{\epsilon}
\newc{\gev}{\,GeV}
\newc{\eq}[1]{(\ref{eq:#1})}
\newc{\eqs}[2]{(\ref{eq:#1},\ref{eq:#2})}
\newc{\etal}{{\it et al.}\ }
\newc{\ibid}{{\it ibid}.}
\newc{\eg}{{\it e.g.}\ }
\newc{\ie}{{\it i.e.}\ }
\newc{\nonum}{\nonumber}
\newc{\lab}[1]{\label{eq:#1}}
\newc{\dpr}[2]{({#1}\cdot{#2})}
\newc{\lsimeq}{\stackrel{<}{\sim}}
\newc{\gsimeq}{\stackrel{>}{\sim}}
\newc{\half}{\frac{1}{2}}

\newc{\rpvs}{{\not R_p}}
\newc{\rp}{{${R_p}$}}
\newc{\kap}{\kappa}
\newc{\ptmiss}{/ \hskip-7pt p_T}
\newc{\mgut}{M_U}
\newc{\wt}{\widetilde}
\newc{\gl}{\wt g}
\newc{\mgl}{m_{\gl}}
\newc{\cnone}{\wt\chi^0_1}
\newc{\cpmone}{\wt \chi^{\pm}_1}
\newc{\mcpone}{m_{\cpone}}
\newc{\mcpmone}{m_{\cpmone}}
\newc{\mcnone}{m_{\cnone}}
\newc{\bit}{\begin{itemize}}
\newc{\eit}{\end{itemize}}
\newc{\delgs}{\delta_{GS}}
\newc{\mth}{m_{3/2}}
\newc{\bea}{\begin{eqnarray}}   \newc{\eea}{\end{eqnarray}}
\newc{\baa}{\begin{array}}      \newc{\eaa}{\end{array}}
\newc{\thetaw}{\theta_W}
\newc{\call}{{\cal L}}
\newc{\mplanck}{M_{\rm P}}
\newc{\dmchi}{\Delta m_{\tilde\chi}}
\newc{\cpone}{\wt \chi^+_1}
\newc{\cmone}{\wt \chi^-_1}
\newc{\gam}{\gamma}
\newc{\cntwo}{\wt\chi^0_2}
\newc{\lampp}{\lam^{\prime\prime}}
\newc{\wtil}{\widetilde}
\newc{\glsp}{$\wtil g$-LSP}
\newc{\mev}{~{\rm MeV}}
\newc{\msq}{m_{\squark}}
\newc{\anti}{\overline}
\newc{\mtil}{\widetilde m}
\newc{\mt}{m_t}
\newc{\mw}{m_W}
\newc{\wbp}{W^+}
\newc{\rts}{\sqrt{s}}
\newc{\wm}{W^-}
\newc{\fbi}{~{\rm fb}^{-1}}
\newc{\chitil}{\wt\chi}
\newc{\cmtwo}{\wt \chi^-_2}
\newc{\cnthree}{\wt\chi^0_3}
\newc{\cnfour}{\wt\chi^0_4}
\newc{\pbi}{~{\rm pb}^{-1}}
\newc{\pb}{~{\rm pb}}
\newc{\etmiss}{/ \hskip-7pt E_T}
\newc{\tanb}{\tan\beta}
\newc{\tb}{\tan\beta}
\newc{\mhalf}{m_{1/2}}
\newc{\mcntwo}{m_{\cntwo}}
\newc{\mcpmtwo}{m_{\cpmtwo}}
\newc{\mcnthree}{m_{\cnthree}}
\newc{\mcnfour}{m_{\cnfour}}
\newc{\cpmtwo}{\wt \chi^{\pm}_2}
\newc{\vev}[1]{{\left\langle #1\right\rangle}}
\newc{\mtilq}{m_{\tilde q}}
\newc{\mtill}{m_{\tilde\ell}}
\newc{\cost}{\cos{\theta_{\tilde t}}}
\newc{\lamepd}{\lam'_{131}}
\def\lapp{\mathrel{\rlap{\raise.5ex\hbox{$<$}}
                    {\lower.5ex\hbox{$\sim$}}}}
\def\gapp{\mathrel{\rlap{\raise.5ex\hbox{$>$}}
                    {\lower.5ex\hbox{$\sim$}}}}
\begin{document}
\begin{titlepage}
\begin{flushright}
TIFR/HEP/03-04\\
{hep-ph/yymmddd}
\end{flushright}
\vskip .6cm
\begin{center}
{\Large\bf  Probing R-parity violating models of neutrino mass at 
the Tevatron via top Squark decays
}\\[1.00cm]
{\large Siba Prasad Das$^{a,}$,\footnote{\it spdas@juphys.ernet.in}
Amitava Datta$^{a,}$} \footnote{\it adatta@juphys.ernet.in}
{and}
{\large Monoranjan Guchait$^{b,}$} \footnote{\it 
guchait@tifr.res.in}
{\\[0.3 cm]}
{\it $^a$ Department of Physics, Jadavpur University,
Calcutta-700032,India}\\[0.3cm]
{\it $^b$ 
Department of High Energy Physics, Tata Institute of
Fundamental Research\\
Homi Bhabha Road, Mumbai-400005, India.}\\[0.3cm]

\end{center}
\vspace{.1cm}

\begin{abstract}
{\noindent\normalsize

We have estimated the limiting branching ratio of 
the R-parity violating (RPV) decay of the lighter top squark, 
$\tilde t_1 \ar l^+ d$ ($l=e$ or $\mu $ and d is a down type 
quark of any flavor), as a function of top squark mass($\MST$) for 
an observable signal in the di-lepton plus di-jet channel at 
the Tevatron RUN-II experiment with 2~fb$^{-1}$ luminosity.
Our simulations indicate that the lepton number violating nature of the 
underlying decay dynamics can be confirmed via the reconstruction 
of $\MST$.
The above decay is interesting in the context of
RPV models of neutrino  mass where the RPV couplings ($\lambda'_{i3j}$)
driving  the above decay are constrained to be small
( $\lsim 10^{-3} - 10^{-4} $). If $\tilde t_1$ is the next 
lightest super particle - a theoretically well motivated scenario -
then the RPV decay can naturally compete with the R-parity 
conserving (RPC) modes which also have suppressed widths.
The model independent limiting BR can delineate the 
parameter space in specific supersymmetric models, where the dominating 
RPV decay is observable and predict the minimum 
magnitude of the RPV coupling that will be sensitive to Run-II data. We
have found it to be in the same ballpark value required 
by models of neutrino mass, for a wide range of $\MST$. A comprehensive
future strategy for linking top squark decays with  models of 
neutrino mass
is sketched.
}
\end{abstract}

\end{titlepage}

\textheight=8.9in

\section*{1.~Introduction}

The Minimal Supersymmetric Standard Model (MSSM)\cite{susy} is a  
well motivated extension of the Standard Model(SM), which is free from 
several shortcomings of the latter. As of now,  there is 
no experimental  evidence either in favor of or against  it.
Unfortunately, the mechanism of supersymmetry(SUSY) breaking is not 
known yet, although several interesting suggestions exist\cite{susy}.
As a result there is no guideline for predicting  
the mass splitting between a SM particle and its superpartner 
and consequently, there is no information  about the range of 
superparticle (sparticle) masses from 
the theoretical point of view.  There are some experimental
lower bounds from unsuccessful collider searches at 
LEP~\cite{lepbound} and Tevatron Run-I~\cite{xsusy,work}.

Currently the Run-II of the Tevatron (referred to hereafter as Run-II) 
is in progress. It is expected to deliver an integrated 
luminosity of at least 2 fb$^{-1}$ per experiment at 2 TeV centre of 
mass energy, which is more than one order of magnitude larger than the 
acquired luminosity in Run-I with center of mass energy 1.8 TeV.
However, in view of the existing limits on the masses of the strongly 
interacting  sparticles (squarks and gluinos) \cite{lepbound,xsusy} 
and the rather marginal increase in  centre of mass energy, most of 
the unexplored parameter space in this sector is  likely  to be beyond 
the kinematic reach of Run-II as well. Yet this is the only currently 
available machine for direct SUSY searches until the LHC starts.

In view of this, the top squark (the superpartner of the top quark) is 
somewhat special. It may be lighter than the other squarks and gluinos 
due to several reasons. Firstly, the large top Yukawa coupling which 
controls the evolution of the soft-supersymmetry breaking masses of 
the left and right handed top squarks, $\tilde t_L, \tilde t_R$, via 
the  renormalization group(RG) equations, tends to reduce these masses 
\cite{susy}. Moreover, because of the large top quark mass, the two 
weak states  $\tilde t_L, \tilde t_R$ may mix very strongly leading 
to a  relatively large splitting between the two physical mass 
eigenstates $\tilde t_1, \tilde t_2~$~\cite{stopmix} (in our notation 
$m_{\tilde t_2} > m_{\tilde t_1}$). Interestingly, the mass of the 
lighter states $\tilde t_1$ may be even below the top mass. In fact, 
it is quite conceivable that in certain region of SUSY parameter space 
it happens to be the next lightest 
supersymmetric particle (NLSP), the lightest neutralino $\tilde\chi_1^0$
being  the lightest supersymmetric particle (LSP) by assumption in 
most R-parity($R_p$) conserving  models. It is, therefore, very
important to fix up the strategies for isolating the top squark signal 
for all conceivable decay modes. Yet another motivation to look for a 
light top squark is that it seems to be preferred by electroweak 
baryogenesis ~\cite{baryosusy}.

In many studies the MSSM is assumed to be a $\R_p$ conserving(RPC) 
theory. The $R_p$ is a discrete symmetry imposed on the MSSM to 
avoid the lepton and baryon number violating interactions in the 
Lagrangian which lead to rapid proton decay. If, however, either 
lepton or baryon number violation is allowed, such catastrophic 
decays can be avoided. This can be achieved by imposing either 
the so called baryon parity  or the lepton parity \cite{bp,lp} 
conservation. The resulting theory,  called R-parity violating (RPV) 
MSSM \cite{dreiner}, is phenomenologically attractive since it has 
many novel predictions. From the theoretical point of view  
both RPC and RPV versions of the MSSM are on equal footing since
both require additional discrete symmetries beyond the gauge symmetry. 

The SUSY signatures are determined by whether R-parity is assumed 
to be conserved or not. Conservation of R-parity implies that all 
SUSY decay chains end up in the $\LSP$, which is stable and escapes 
the detector. Thus  a typical SUSY signature is always accompanied by 
some amount of missing transverse 
energy ($\met$). The R-parity violating interactions ~\cite{dreiner,france} 
on the other hand would allow the LSP, which is not necessarily the 
$\LSP$, to decay into SM particles leading to distinct signals
\cite{rohini}. Yet another type of signal is the  direct decay of 
sfermions, into two SM particles \cite{france}. In view of 
the large production cross-section of $\t1  \t1^*$ pairs, the class of 
lepton number violating decays generically denoted by, 
\be
\tilde t_1 \ar l_i^+ d_j
\label{rpvmode}
\ee
is an attractive channel for searching RPV interactions at Run-II. Such 
decays are driven by the $\lambda'_{i3j}$ $L_i~ Q_3~{\bar D}_j^{c}$
term in the superpotential where L, Q and D are respectively  the 
lepton doublet, quark doublet and down singlet type 
superfield and $i,j$ are generation indices.

In recent times RPV models have attracted special attention as they can 
provide  viable models of neutrino mass ($m_{\nu}$). The basic mechanism
has been known for a long time\cite{numassold}. The interest in this model 
was revived after the atmospheric\cite{atmos} and solar 
neutrino\cite{solar} experiments confirmed that the
neutrinos are not massless. Parameters of these
models have been constrained by many groups using  neutrino data 
\cite{numassnew,abada}. The actual set of parameters ( bilinear and trilinear
terms \cite{dreiner} 
in the RPV sector of the superpotential) and their precise
magnitudes required to explain the neutrino data is model 
dependent. However some couplings belonging to the class 
$\lambda'_{i3j}$ are important ingredients of model building. Considering 
a variety of models it has been shown, e.g., in~\cite{abada}, that 
the important couplings $\lambda'_{i33}$, for all $i$, turn out to be 
generically small 
( $\lsim 10^{-3} - 10^{-4} $, depending on the magnitude of the 
soft breaking parameters in the RPC sector).
This is certainly much stronger than the 
constraints obtained prior to the neutrino data\cite{gbherbi}. Thus
RPV decays of the top squark driven by these
couplings may provide an avenue for probing the models 
of $\nu$-mass ~\cite{asesh,valle,subhendu} at colliders.

In this paper we focus  our attention  on two interrelated topics :

\begin{enumerate}

\item  The viability of observing direct top squark decays through the
lepton number violating channels, Eq.(\ref{rpvmode}),
at the upgraded Tevatron collider in a model independent way 
using the event generator {\tt PYTHIA} ~\cite{spythia}. This is
an issue important in its own right irrespective of models of $m_{\nu}$. 
\item
The implications  of observation/non-observation of this decay channel 
for models of neutrino mass.
\end{enumerate}

The collider signatures, however, crucially depends on whether the top squark
is the NLSP or not. If the top squark is not the NLSP and the RPV couplings 
are as small as that required by the models of the neutrino mass, it  
would dominantly decay via the RPC 2-body mode with nearly 100\% BR, 
~\cite{hikasa,2bdk}
\be
\tilde t_1 \ar b \tilde\chi_1^+
\label{2dk}
\ee

where  $\chplus$ is the lighter chargino, 
or, if the above mode is not kinematically allowed, via the 3-body modes 
\cite{threebody}, 
\be
\tilde t_1 \ar b \ell \tilde\nu ,~ b \tilde\ell \nu ,~ bW \LSP
\label{3bdk}
\ee
where $\tilde\ell$ and $\tilde\nu$  are respectively the slepton and 
the sneutrino assumed to be lighter than $\t1$.
The decay of the LSP would then be the only signature of RPV interactions.  
Whether the magnitude of the underlying RPV coupling is indeed in the 
right ballpark or not, can be tested in principle, e.g.,  by 
measuring the width of 
the LSP, which may not be  an easy task at least in the context of 
Run-II.   Of course neutrino masses are generated by specific RPV parameters, 
e.g., by the lepton number violating trilinear couplings $\lambda'_{i33}$, 
where $i$ is the lepton generation index, if the bilinear couplings 
happen to be small.
Thus the decay patterns of the LSP  may give some  
circumstantial evidence in favor of/against  models of neutrino mass. 
For example, if
$\LSP$ is assumed to be the LSP, then $\chplus_1 \chminus_1 $ 
and   $\chplus_1 \LSP $ production
followed by decay chains involving the decays 
\be
\LSP  \ar \nu_{\mu} b \bar b  ,~ \nu_{\tau} b \bar b
\label{lsp3bdk}
\ee
is indicative of  an underlying  model of neutrino mass \cite{barger}. 
In ref.~\cite{barger} the prospect of observing this signal at 
Run-II was 
studied. It was concluded that
this signature can be probed up to $\mhf$ = 230~GeV(320 GeV) with an 
integrated luminosity of 2 $fb^{-1}$ ( 30 $fb^{-1}$ ). Here $\mhf$ 
is the common gaugino mass at the GUT scale. 
It is, however, worth noting  that since the signal  has
missing energy carried by the neutrinos, it can be mimicked 
even if R-parity is conserved. 
For example, the decay  $\LSPsecond 
\ar \LSP b \bar b $, which may have a large BR if one of the bottom 
squark mass eigenstates happens to be lighter at large 
$\tb$, has collider signatures very similar to the decay 
of Eq.(\ref{lsp3bdk}).
Moreover, the lepton number violating nature of the 
underlying interaction 
is not obvious since the neutrinos escape the detector. The 
possibility of probing RPV models of neutrino mass through neutralino 
decays has also been considered in \cite{rpvtev}.

The situation is totally different if the lightest neutralino and  
the top squark happen to be the LSP and the NLSP respectively,
a theoretically well motivated scenario for reasons discussed above. In this
scenario the main RPC decay channels occur via the flavor changing neutral 
current decay mode~\cite{hikasa},
\begin{equation} 
\tilde t_1 \ar c \tilde\chi_1^0
\label{loopdk}
\end{equation}
and via 4-body decay modes with a b-quark, $\LSP$ and two massless 
fermions~\cite{boehm},
\begin{equation} 
\tilde t_1 \ar b \tilde\chi_1^0 f \bar f' 
\label{4bdk}
\end{equation}
( $f$ and $\bar f'$  being a quark-antiquark or $l$-$\bar\nu_l$ pair ) 
which eventually lead to RPV 
signals due to the LSP decay. Here the key point is that the above channels 
have widths suppressed due to
natural reasons and can very well compete with each other~\cite{boehm} or 
with the RPV mode even, if the coupling 
$\lambda'_{i3j}$  is $\sim 10^{-3} - 10 ^{-4}$. 
As we shall see in a later section such competitions occur naturally over 
a large region of the MSSM parameter space. In fact if the above coupling is 
much larger then the direct lepton number violating decay mode
Eq.(\ref{rpvmode}) will occur with 100\% BR's. The coexistence of the
direct lepton number violating decay mode as well as RPC decay modes followed
by the LSP decay, is a hallmark of RPV models of neutrino mass. Moreover 
this signal  is attractive due to  the 
large production cross section of  top squark pairs~($\t1 \t1^*$). In 
addition, if a signal is observed then the underlying lepton number violating 
interaction can be revealed easily by reconstructing the top 
squark mass as will be illustrated in a later section. 
The signatures of this decay at the Tevatron was first discussed 
in~\cite{dpmg} in the context of  high $Q^2$ events at HERA ~\cite{Q2}.

Obviously the presence of competing channels 
may complicate the search for the top squark. For example, if
each of the competing modes has BRs substantially smaller than 100\% all of
them may be below the observable level in spite of large production rate
of  $\t1 \t1^*$  pairs. A complete 
discussion is not possible without full simulations 
of all possible signals which is beyond the scope of this paper. We shall 
concentrate on the first task, namely to estimate the 
minimum BR of top squark decay in RPV channel, Eq.(\ref{rpvmode}), 
required for the observation of the signal 
at Run-II. This will be done in a subsequent section using the MC event 
generator {\tt PYTHIA} ~\cite{spythia}.

In a recent paper it has been shown~\cite{subhendu} that the data from 
Run-I of the Tevatron already restrict the BR of the decay, 
Eq.(\ref{rpvmode}) to values significantly smaller than 100\% 
in a model independent way for a range of top squark
mass ( see Fig.~3 of \cite{subhendu}). 
Assuming specific model parameters, which fixes the BRs of all the competing
channels,  this BR exclusion can be translated into upper bounds on the 
corresponding RPV coupling ( see Fig.~4 of \cite{subhendu}). 
It was found for the first time, albeit for small values of the top 
squark mass and rather limited regions of the  MSSM parameter space, 
that the Run-I data were 
indeed  sensitive to magnitudes of these couplings relevant for models 
of neutrino mass. Since the accumulated luminosity
of Run-II is more than an order of magnitude larger, we feel encouraged 
to investigate the feasibility of obtaining similar constraints over a  
much larger region of the MSSM parameter space. It may be recalled that
in the past Tevatron di-lepton data was also used to constrain the squark
and gluino masses in the context of RPV SUSY model~\cite{dilepton}. 

The possibility of probing the RPV models of neutrino mass via top 
squark decays was also suggested in Ref.\cite{asesh,valle}. These works, 
however, differ from ours in several ways.  In~\cite{asesh}, which was 
the first attempt to confront Run-I data with models of neutrino mass, 
it was claimed that for values of RPV coupling favored by models of 
neutrino mass, the RPV decay of the top squark dominates over the loop 
decay for $\MST<$ 150 GeV. In the absence of
4-body decays this statement is correct for low $\tb$ only.
For high $\tb$ the loop decay can overwhelm  the RPV mode. 
For low $\tb$, on the other hand, the 4-body decay, Eq.(\ref{4bdk}), 
not considered in~\cite{asesh}, may  dominate over the RPV decay
if $\lambda' \approx 10^{-3}- 10^{-4}$. 
The SU(2) gaugino mass $M_2$ also plays an important role in determining 
the relative 
strengths of these competing modes. All these issues will be 
addressed in great details in a subsequent section.
In Ref.~\cite{valle} the RPV mode, Eq.(\ref{rpvmode}), the loop decay
Eq.(\ref{loopdk}),  and several
3-body modes (including those in Eq.(\ref{3bdk})), were assumed to be 
the competing channels.  However, no detailed simulation was carried out 
to estimate  the sensitivity  of the data to RPV couplings.

We have organized the paper as follows. In Sec. 2, we shall describe 
our road map for obtaining  a comprehensive  search strategy 
for top squark and its consequences for models of $m_\nu$ and briefly review 
the current status of top squark search, especially when it happens to be 
the NLSP. In Sec. 3, the details of the simulation leading to 
  model independent limiting values of the BR ($ \tilde t_1 \ar l^+ d_j $)
, where $l=e$ or $\mu$, sensitive to  Run-II data will be presented
as a function of $\MST$. 
In Sec. 4, we 
use the results of Sec. 3 to obtain upper limits on RPV couplings
in specific models and to understand the systematics of the parameter
space (i.e. delineating  the regions where some of the competing modes
dominate or several of them may coexist).
We summarize our results in Sec. 5.

\section*{2.~The road map for linking top squark search with models of
neutrino mass}

Our  first task is to  assess the viability of observing
the RPV decay, Eq.(\ref{rpvmode}) at Run-II. For simplicity we shall
as usual assume that the RPV couplings
are hierarchical, i.e., one coupling of the type $\lambda'_{13j}$
dominates over the others. 

For the 
sake of definiteness our simulations will be restricted to the mode
$\tilde t_1 \ar e^+ d_j $. 
This decay is triggered 
by the trilinear RPV coupling  $\lambda'_{13j}$ $L_e Q_3{\bar D}_j^{c}$ 
term in the superpotential \cite{dreiner}, where $j$ = 1-3 is a 
generation index for down type quarks. 
In order to make our analysis as general as 
possible, we have not
employed any 
particular jet tagging so that the  conclusions are approximately valid 
for any $j$.{\footnote {
The most important ingredients  of RPV models of $m_\nu$ 
are $\lambda'_{i33}$  all of which 
are constrained to be  $\lsim 10^{-3} - 10^{-4} $ ( See, e.g, ~\cite{abada}
). However, for $d_j$=b, b-tagging can be efficiently employed to improve
the signal/background ratio and our conservative conclusions may be further strengthened. 
( see Sec.~3 for further comments).}}

Our conclusions are also approximately valid for the coupling  
$\lambda'_{23j}$ $L_\mu Q_3{\bar D}_j^{c}$. A small difference 
may arise due to the difference in the detection efficiencies for 
$e$ and $\mu$. However, since the leptons 
are highly central the difference is rather 
marginal. Our conclusions cannot be applied to the signal from  the 
$\lambda'_{33j}$ $L_\tau Q_3{\bar D}_j^{c}$ term which requires a fresh 
simulation taking into account $\tau$ detection efficiency.
We, however, feel that the simplest signal arising from the class 
of decays in 
Eq.(\ref{rpvmode}) will be sufficiently informative for the first analysis 
using an event generator.

A systematic search strategy for the top squark  or, in the absence of a 
signal, a comprehensive limit on $\MST$ in RPV MSSM, therefore 
depends on several steps. The first step is to estimate the model 
independent minimum value 
$\sigma(p\bar p \ar \tilde t_1 \tilde t_1^{*})*(\epsilon_{br})^2$ for an 
observable signal as a function of $\MST$, where
$\epsilon_{br}$=BR($\tilde t_1 \to e^{+} d_j$).  Using the well-known 
formula for 
$\sigma(p\bar p \ar \tilde t_1$
$\tilde t_1^{*})$, which is available up to  next to leading 
order (NLO) \cite{spira}, this bound can be translated into a 
lower limit 
on observable BR. Rather low values of 
$\epsilon_{br}$ can be probed for a range of $\MST$    
and $\MST$ may be reconstructed with reasonable accuracy
at Run-II, as we shall see in the next section.

The observation of the RPV signal alone, though a stupendous achievement 
in its own right, will shed little light on models of neutrino mass($m_\nu$). 
As discussed in the  introduction the simultaneous observation of the 
signals arising from the RPC decays in Eq.(\ref{loopdk}) or Eq.(\ref{4bdk}), 
followed by $\LSP$ decay may strongly hint
in favor of these models. The observability of these signals depend on
two 
factors,  a) the BRs of the decays involved and b) the acceptance efficiency 
of the cuts in distinguishing the signal from the background.

Assuming that the dominant RPV decay mode 
of $\LSP$ in RPV models of $m_\nu$ is 
$\LSP  \ar b \bar b \nu_i $, $i$=1,2,3,  the signal 
resulting from the loop decay (Eq.(~\ref{loopdk})) is 
\begin{equation}
\tilde t_1 \ar c \tilde\chi_1^0 \ar c b \bar b \nu_i 
\label{loop_lspdk}
\end{equation}

Therefore,$\t1  \t1^*$ pair production is signaled by  
jets+$\met$ with 4 b-jets. Similarly the 4-body decay, Eq.(\ref{4bdk}) 
would cascade into  
\begin{equation} 
\tilde t_1 \ar b \tilde\chi_1^0 f \bar f' \ar b b \bar b \nu_i f \bar f' 
\label{4bd_lspdk}
\end{equation}

An excess of $\ell$+jets+$\met$, 2$\ell$+jets+$\met$ or jets+$\met$ events
including several b-jets
would, therefore, indicate $\t1  \t1^*$ pair production in the framework of
RPV SUSY model. The above
signals are very similar to the ones discussed in ~\cite{barger} although 
the signal from the decay chain in  Eq.(\ref{4bd_lspdk}) may have even
more b jets. From the results of ~\cite{barger} one has reasons to 
be optimistic that the large number of b jets would provide a visible signal
if b-tagging is really efficient($\approx 50\% $).

Full simulations of the above two signals, which is 
beyond the scope of this paper, would lead to the estimated 
minimum BR of the loop decay  and the 4-body decay required 
for observable signals at Run-II as a functions of $\MST$. These 
along with the minimum BR for observable RPV signal ( estimated
in the next section in detail) will provide the basis for a model
independent approach to top squark 
search at Run-II in the context of RPV theory of $m_\nu$.

If the signal is seen in the RPV channel as well as in one or both of the 
competing channels, one can try to identify the allowed parameter space
using the limiting BR and the reconstructed $\MST$.  Since the 
estimates of the limiting BR corresponding to the signals 
in Eq.(\ref{loop_lspdk}) and  Eq.(\ref{4bd_lspdk}) are not 
available at the moment, a complete job can 
not be done.  However the BR of the RPV decays will
be discussed in details in Sec. 4 with an aim to understand
the systematics of the MSSM parameter space vis-a-vis
these decays.  Outlines of a future comprehensive  programme
for linking $\t1$ decay signals with models of $m_\nu$
will be  sketched with illustrative example in Sec. 4.

The other important issue is the  prospect of unambiguously excluding 
a range of $\MST$ if no signal is seen. Here one encounters the complications 
due to possible presence of three competing  decay modes  
in a large parameter space. In fact the current mass limits 
on $\MST$ in both RPC and RPV models are  also not free from ambiguities.

The phenomenology of top squark search Tevatron experiments in different 
decay channels have been studied extensively in both RPV 
~\cite{asesh,valle,subhendu} and RPC \cite{sender,mambrini,lykken,admgspd}
models. The  unsuccessful search for the top squark at LEP and Tevatron 
Run-I experiments in both RPC ~\cite{rpc20} and RPV MSSM~\cite{rpv21,cdfstop}
have yielded important bounds. Here we shall focus on the scenario when 
top squark is the NLSP.

It is to be noted that the most stringent limits in RPV as well as 
RPC models have often been derived by employing the model dependent 
assumption that the top squark decays into  a particular channel     
with 100\% BR. For example, the most stringent bound in the context 
of RPC MSSM comes from Tevatron Run-I  experiments which puts a lower 
limit on lighter top squark  mass $m_{\tilde t_1} \ge $ 119 GeV for 
$m_{\tilde\chi^0_1}=$40 GeV. The limit becomes weaker 
for higher value of $m_{\tilde\chi^0_1}$, e.g.,  $m_{\tilde t_1} \ge $ 
102 GeV for $m_{\tilde\chi^0_1}=$50 GeV~\cite{rpc20}. In deriving these 
limits, it was assumed that the loop induced, decay Eq.(\ref{loopdk})
~\cite{hikasa}, 
occurs with 100\% BR. Apparently this  assumption is  
valid in a wide class of models if the $\tilde t_1$ state happens to 
be the NLSP. Since the production cross section of top squark pairs is 
dominantly via QCD and depends on its mass only, the above limits from 
Tevatron, therefore, seem to be  fairly model independent, except for 
the dependence on $m_{\tilde\chi_1^0}$, which influences the 
efficiency of the kinematical cuts.

However, as has been shown in \cite{boehm}, even if the top squark is 
the NLSP, its 4-body decay, Eq.(\ref{4bdk}),
may indeed compete with the above loop decay or may even overwhelm 
it in some region of parameter space. The above limits, therefore, 
require revision and new signal via the 4-body decay channel 
should be looked for\cite{admgspd}.

The most recent limit on the top squark  mass ( $\MST \gsim$ 122 GeV ) 
in the RPV MSSM comes from the CDF collaboration \cite{cdfstop} in the 
decay channel 
\be
\tilde t_1 \ar \tau^+ + b
\ee
This limit is also  derived on the basis of the above model dependent 
assumption, namely, the decay channel in question has a BR 
of 100\%. However, even if the RPV coupling 
involved ($\lambda'_{333}$ ) 
is assumed to be the most dominant one, the mode may have 
a BR significantly smaller than 100 \%. This may happen in various 
regions of the MSSM parameter space simply due to the competition 
among this decay mode and several RPC modes of top squark, since the latter
couplings are invariably present in the theory irrespective of 
the choice of the RPV sector. As discussed in the introduction the 
competition is of special interest, if the top squark is the NLSP and 
RPV couplings have
strengths relevant for the models of neutrino mass \cite{numassnew,abada}. 
In \cite{subhendu}
on the other hand the possibility of competition among different decay 
channels were considered. The
mass limits obtained in \cite{subhendu} were naturally dependent on 
BR($\tilde t_1 \ar e \bar d $ )= $\epsilon_{br}$. For example it was 
found that $\MST \gsim$ 200(165) GeV 
for $\epsilon_{br}$=1(0.5).

If no RPV signal is seen at Run-II, any particular $\MST$ can not be 
excluded in a model
independent way. Only the regions of the MSSM parameter space where 
the BR of at least one of the three competing decay modes is above the 
observable limit will be ruled out. On the other hand
one can also identify  the difficult regions of the MSSM 
parameter space, in the 
context of Run-II , where all three decay modes have low BRs. 
The stop search at LHC may focus on these regions. In the 
difficult regions  the RPV signals from chargino/neutralino production 
followed by  $\LSP$ decay ~\cite{barger} appear to be the only possibility 
of probing models of $m_\nu$ at Run-II . Thus the top squark decay and 
the signal of ~\cite{barger} are essentially complementary in nature.

It should also be noted that the above top squark decay signals 
are important only if the top squark happens to be  the NLSP,  a 
scenario  theoretically very well motivated but not inevitable. 
The signal of \cite{barger} on the other hand requires the lighter chargino
to be heavier than the $\LSP$ which is not necessary in RPV models, 
unless gaugino mass unification~\cite{susy} is assumed. Thus 
either of the above two signals, Eq.(\ref{rpvmode}) or Eq.(\ref{lsp3bdk}), 
or both may be helpful for probing RPV signals depending on the MSSM 
parameter space of interest.

The limit on the RPV  BR in turn can 
be converted into upper limits on $\lambda'$ in specific models with 
several competing channels. We shall demonstrate in Sec. 4 that for 
a wide choice of model parameters magnitudes of $\lambda'$ relevant 
for models of $m_\nu$ are expected to be sensitive to the data.

Once the  LHC is in operation the signal size as well as the 
ability to probe smaller BR are expected to increase 
dramatically. The task of reconstructing $\MST$, and 
delineating the allowed/disfavored regions of the parameter 
space in specific models will be much easier.
The programme for a comprehensive top squark search will 
certainly take some time. Yet, it is gratifying 
to note that a systematic, largely  model independent strategy for 
top squark search in models of $m_\nu$  is quite possible in a not too distant future.

\section*{3.~The Limiting Values of 
BR($\tilde t_1 \ar e^{+}  d)$ for  Observable Signals at  Run-II }

In hadron colliders, top squark pairs are produced via 
gluon-gluon fusion and
quark-antiquark annihilation,
\be
gg,q \bar q \ar \tilde t_1 \tilde t_1^*
\ee
The production cross section depends only on the mass of 
$\tilde t_1$ without any dependence on the mixing angle in the 
top squark sector, since it is a pure QCD process~\cite{kane}.
The total pair production cross section at the Tevatron for 
$\sqrt s=$ 2 TeV is $\simeq$ 15-0.3 pb which is 40\% larger 
than the cross section for $\sqrt s=$ 1.8 TeV, for the range 
of $m_{\tilde t_1}\sim$ 100-200. The QCD corrections enhance 
this cross section by $\sim$30\% over most of SUSY parameter 
space accessible at Tevatron~\cite{spira}.

We investigate the signal of top squark pair production in the 
channel $e^{+} e^{-}$ plus two or more jets, assuming that 
both the top squark decays via a single  RPV coupling $\lambda_{13j}$,
\be
\tilde t_1 \ar e^{+} +d  \\ \nonumber 
~ ; ~\tilde t_1^* \ar e^{-} + \bar d, 
\label{stopsig}
\ee
where we have suppressed the generation index of the d type quark since 
we have not employed any specific flavor tagging.

The leading SM backgrounds corresponding to the signal with 
{\it opposite sign di-electron ( OSDE)  plus 2 or more jets}
are the following:
\begin{enumerate}
\item[a.]
Drell-Yan process via $q \bar q' \to e^+ e^-$.

\item[b.]
$W$ boson pair production, $q \bar q' \to W W$, where both the W 
decay leptonically, $W \to e \nu_e$. Note that we also consider 
W decays to $\tau$ leptons which may decay to electrons.

\item[c.] 
$q \bar q' \to W Z$, where W decays hadronically and $Z$ decays  
leptonically.

\item[d.]
$Z$ boson pair production also  leads to the 
same final state : $q \bar q' \to ZZ \to (q\bar q')(e e)$. 

\item[e.] Top quark pair production, $q \bar q, g g \to t \bar t$ , 
where both the top quarks decay semi-leptonically via $W$,
$ t \to b e \nu_e$.

\item[f.]
Single top quark production, $q \bar q' \to t b$, where one 
lepton comes from top quark and the other comes from $b$-quark 
decay.    

\end{enumerate}
In processes a) - d) additional jets come from initial/final state 
QCD radiation(I/FSR). We have analyzed the signal and background processes 
using the {\tt PYTHIA} (v6.206) event generator~\cite{spythia}. 
We generate signal events in the di-electron + jets channel 
forcing $\t1$ to decay, $\t1 \to e + q$ with 100\% branching 
ratio switching off all other allowed decay modes of $\t1$ in $\tt PYTHIA$. 

In our calculation we set the renormalization and factorization 
scale to $Q^2=\hat s$ and CTEQ3L~\cite{cteq} for the parton 
distribution functions. For the jet reconstruction we use the 
routine {\tt PYCELL} in {\tt PYTHIA} ~\cite{spythia}. We selected 
events in the hadronic and electromagnetic calorimeter cells in 
pseudorapidity ($\eta$) and azimuthal angle($\phi$) of size 
$\Delta \eta \times \Delta \phi=0.1 \times 0.1$. Cells with 
$E_T > $ 1 GeV are taken as initial seeds to form calorimetric 
towers. Jets are reconstructed with cone radius 0.5 and only those 
are accepted which has transverse energy $E_T >$ 8 GeV and are
smeared by 0.5$\times \sqrt {E_T}$. 
We selected events applying the following set of cuts.

\begin{enumerate}
\item
Leptons, required to be of opposite charges and of same flavor, 
are selected with $p_T^\ell >$10 GeV and $|\eta_\ell|<$2.5

\item
Number of jets are required to be $n_j \ge 2$, where, 
jets are  selected if $ E_T^j > $ 15 GeV, $|\eta_j|<$3.0.  Isolation 
between any two jets are ensured by demanding $\Delta R(j,j)>$ 0.5, where 
$\Delta R = \sqrt{\Delta\phi^2 + \Delta\eta^2}$.

\item
Electrons and jets are assumed to be isolated, if $\Delta R(\ell,j)>$ 0.5. 

\item
Events with di-electron invariant mass between 
80 GeV$< M_{\ell\ell}<$ 100 GeV 
and $ M_{\ell\ell}<$ 10 GeV are not accepted.

\item
Azimuthal angle between two leptons are required to be $\phi(\ell\ell) 
< 150$ degree. 

\item
Events are vetoed out for $p{\!\!\!/}_T >$ 25 GeV. 

\item
The total visible energy of any event are required to be, 
$S_T > 350$ GeV, 
where $S_T = H_T^\ell + H_T^j$; $H_T^{\ell(j)}$ = scalar sum of 
transverse energy of all leptons(jets). 

\item
We constructed two lepton-jet invariant masses considering all possible 
combinations of  the final state particles.
Finally  we select only that combination  in which the difference 
between two is minimum provided 
$|m(\ell_1 j_1) - m(\ell_2 j_2)| <$ 20 GeV.
\end{enumerate} 

Cuts 1-3 are basically event selection cuts. Cut 4 is used to 
suppress the backgrounds where lepton pair is coming due to 
$Z \to \ell \ell$ decay, where as cut 5 is applied aiming to 
suppress the background due to the Drell-Yan process (a), 
where leptons are mostly back to back in the azimuthal plane. 
Note that the signal is almost free from any missing 
momentum 
{\footnote {Some amount of missing 
momentum may arise if jets or leptons  go undetected due to loss in the 
beam pipe, for very low energies which are  below the detection 
threshold or due to energy momentum mismanagement. In any case, 
the missing momentum  is not so hard.}. Therefore, using cut 6 we 
vetoed out those events which involve large amount of missing momentum. 
The background from $WW$ and $t \bar t$ suffer heavily because of 
this cut. Finally, the cut on total visible energy, $S_T >350$ GeV 
significantly reduces all backgrounds particularly DY, to a negligible 
level, without costing too much in the signal cross section,  except 
for low values of $m_{\t1}$.   

In Table~1, we summarize our results for all backgrounds. The second 
column contains the raw production cross sections corresponding to each 
process. In the 3rd column, we present the number of events 
($N_{1-6}$) surviving after cuts 1-6. The effect of cut 7 is shown 
separately in the 4th column($N_7$). In the 5th column we present the 
acceptance efficiencies for each of the respective processes. 
We notice that the  jet selection cuts are very effective in 
eliminating the backgrounds due to gauge boson pair productions, as  
jets are not very hard in these processes. In the $WW$ case, jets 
mainly arise due to  ISR and  are very soft. We notice 
that in this case the selection efficiency turns out to be at the level of  
$\sim 10^{-4}$ due to the  jet selection cuts. On the other hand, 
in $ZZ$ and $WZ$ case, the lepton pair  comes from $Z$ decay, 
$Z \to e e$, where as the accompanying gauge boson decays hadronically. 
Therefore, although the jet selection cuts are less stringent, 
cut 4 and 5 are very effective.  Finally, the cut on total visible 
transverse energy drastically reduce all background processes 
bringing them to a negligible level. This table clearly shows that 
our signal cross section  is almost background free. The last criterion  
8 is used to reconstruct top squark masses and to reveal the lepton 
number violating nature of the underlying interactions.

\begin{table}[!htb]
\begin{center}
\begin{tabular}{|c|c|c|c|c|c|}
\hline
Process  & Cross section &  $N_{1-6}$ & $N_{7}$ & Efficiency
& Events for \\
 & ( pb) & &  & ($\eps_k$) & ${\cal L}$=2~fb$^{-1}$\\
\hline
WW & 8.06 & 21  & 0  & 0.0 & 0.0
\\
\hline
WZ & 2.34 & 487   & 5 & 1.0$\times 10^{-5} $& 0.047
\\
\hline
ZZ & 1.08  & 1838  & 46 & 9.2$\times 10^{-5}$ & 0.199
\\
\hline
$t \bar t$ & 4.38 & 743  & 30 & 3$\times 10^{-5}$ & 0.263
\\
\hline
$tq$ & 2.39 & 6  & 0 & 0.0 & 0.0 \\
\hline
DY &  2.959$\times 10^4$  &  94 &  1  &  5$\times 10^{-8}$ & 2.959 \\
\hline
\end{tabular}
\vspace*{-2mm}
\end{center}
\caption{Results of the {\tt PYTHIA }\cite{spythia} simulation for
all background processes.
}
\end{table}

In Table~2 which is of the same structure as Table~1 except 
for the last two columns, we show the signal cross sections for 
various top squark masses. It is to be noted that the cut $S_T >$350~GeV 
costs signal cross section heavily ( by about factors of $\sim$ 10-60) 
for lower values of $m_{\t1}$($\lsim$ 80-100~GeV) as leptons and
jets are relatively soft where as for higher $m_{\t1}$ this cut does not 
affect the signal cross section significantly. The signal efficiencies 
vary from $\sim$ 2-30\% for the range of $m_{\t1}$ 100-240 GeV. In 
column 7 we present the significance of the signal for $\epsilon_{br}=$1.  
The last column presents the minimum value of $\epsilon_{br}$ that can be
measured at 5$\sigma$ for an integrated  luminosity of 2~fb$^{-1}$.

\begin{table}[!htb]
\begin{center}
\begin{tabular}{|c|c|c|c|c|c|c|c|}
\hline
$m_{\t1}$  &  $\sigma(p\bar p \ar \tilde t_1$$\tilde t_1^{*})$ & $N_{1-6}$ &  $N_{7}$  & Efficiency
& Events for & S/$\sqrt B $ & limiting BR \\
(GeV) & ( pb) & & & ($\eps_k$) & ${\cal L}$=2~fb$^{-1}$ & & (\%) \\
\hline
80 & 28.09 & 101427 & 1681 & 0.0056 & 314.6 & 168.93 &17.2\\
\hline
100 & 8.59 & 106450 & 5826 & 0.0194 & 333.3 & 178.97& 16.7\\
\hline
120 & 3.18 & 109501 & 14191 & 0.0473 & 300.8  & 161.53 &17.6 \\
\hline
140 & 1.34 & 111196 & 28015 & 0.0933 & 250.0  & 134.26 &19.3 \\
\hline
160 & 0.617 & 111424 & 46985 & 0.1566 & 193.2 & 103.76 & 21.9\\
\hline
180 & 0.304 & 111588 & 68625 & 0.2287 & 139.0& 74.66  & 25.8 \\
\hline
200 & 0.158 & 108769 & 83943 & 0.2798 & 88.41 & 47.47 &32.4  \\
\hline
220 & 0.084 & 106124 & 92213 & 0.3073 & 51.62& 27.72  & 42.5\\
\hline
240 & 0.046 & 103818& 96204 & 0.3206 & 29.49& 15.83  & 56.2\\
\hline
260 & 0.026 & 93410& 89399 & 0.2979 & 15.49& 8.32  & 77.5\\
\hline
\end{tabular}
\vspace*{-2mm}
\end{center}
\caption{Results of the {\tt PYTHIA }\cite{spythia} simulation for the signal 
di-electron plus 2 or more jets due to the top squark pair production at Tevatron  
}
\end{table}

In Fig.\ref{fig_stop}, we show these minimum BRs as a function $\MST$. 
The upper region 
of the solid line can be explored by Run-II for ${\cal L}=$2~fb$^{-1}$ 
luminosity. In the same plane we show the region ( above the dashed line)
which is already excluded at 95\% C.L by Tevatron data~\cite{subhendu}.
Comparing the two regions we find that the improvement in sensitivity is 
by $\sim$ factor of 2-3 for 80 $ \lsim \MST \lsim $ 160. 
For higher top squark masses it is still quite significant. As discussed 
in Sec.2 this is the first step for obtaining a model 
independent search in the framework of RPV MSSM.

The actual limiting BR may be even
smaller as can be seen,  a) by replacing the cross-sections
from {\tt PYTHIA} ( the second column of the table 2) by the corresponding NLO
cross-sections of ~\cite{spira} which are typically
larger by 30\%,  b) if accumulated luminosity
significantly larger than 2$fb^{-1}$ is considered. Our results are therefore
very conservative. More optimistic results can
be easily obtained by dividing the limiting BR in Table~1 by
$(\sigma_{NLO} / \sigma_p)^{1/2}$  $({\cal L_A} / 2fb^{-1})^{1/4}$, where
$\sigma_{NLO}$ is the cross-section in \cite{spira}, $\sigma_p$ is the
{\tt PYTHIA} cross-section 
and ${\cal L_A}$ is the actual luminosity. 

We have not tagged the
flavor of any jet in the final state. We have checked that
for $\MST$=120(180) the overall efficiency,
in Table~2 is reduced to 0.021(0.097) ( including a
b-tagging efficiency of 50\%), if
$d_j $ is identified with a b-quark. This  suppression, however,
will be adequately compensated by strong reduction
in the  backgrounds. For example the Drell-Yan background
will now be non-vanishing mainly due to misidentification of light quark and
gluon jets as b-jets, the probability of which is extremely small.
Assuming the signal to be essentially background free and requiring
10 events as the criterion for discovery, the limiting BR is found to be
27.3\%(41.2\%) for $\MST$=120(180). Due to the uncertainties in cross-section
and ${\cal L_A}$ ( see above) these limiting BR may be smaller. We therefore
feel that the numbers in Table~2 are fairly representative for all
d-type flavors.

\begin{figure}[!htb]
\vspace*{-3.5cm}
\hspace*{-3.0cm}
\mbox{\psfig{file=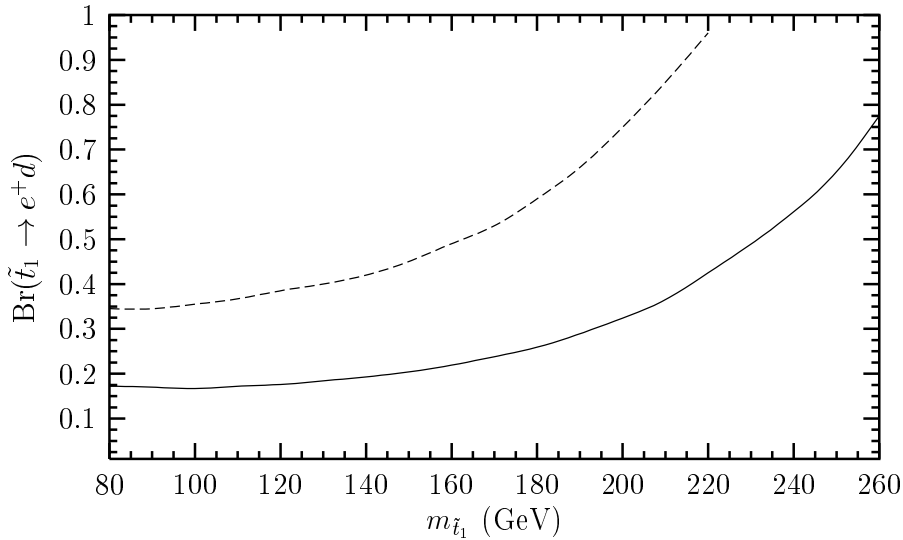,width=20cm}}
\vspace*{-16.7cm}
\caption{The discovery region above the solid line for 
${\cal L}=$2~fb$^{-1}$ at Run-II. The region above the dashed line
is excluded by Tevatron Run-I data~\cite{subhendu}.
}
\label{fig_stop}
\end{figure}

As we mentioned in the previous section, the limiting values of 
$\epsilon_{br}$ can also lead to constraints in the MSSM parameter space
in RPV models of $m_\nu$. We discuss them details in the next section. 
In these models the couplings $\lambda'_{i33}$, $i$=1-3, are the most 
relevant ones in most scenarios. Considering di-leptons of the same 
flavor the BR in Fig.\ref{fig_stop}
may be interpreted as BR ($ \tilde t_1 \ar e b$) or  
BR($ \tilde t_1 \ar \mu b$).

It is expected that the invariant mass of the lepton and jet should show 
up a peak at $\MST$. However, a combinatorial 
problem arises when the decay of a pair of $\t1$ is considered. The last 
kinematic selection 8 is used to reconstruct the top squark mass.  
The correct lepton-jet combination can be separated out by demanding
the difference between any two lepton-jet invariant mass($m_{\ell j}$) 
be the minimum. In Fig.\ref{lep_jj}, we show the lepton-jet invariant mass 
distribution normalized for ${\cal L}$=2~fb$^{-1}$ and with 
$\epsilon_{br}$=1 and for three $\MST$ masses, 100~GeV, 120~GeV and  
140~GeV which are presented by solid, dashed and short dashed lines 
respectively. We have not shown the corresponding distributions for any 
of the backgrounds since
after imposing all cuts those are turn out to be
negligible(see Table~1). As expected, visible peaks at each $\MST$ is 
present which are not expected in any of the backgrounds. 
Therefore, in this channel, the mass of $\t1$ can be determined 
with reasonable accuracy. More importantly the successful reconstruction 
of the top squark mass unambiguously imply the lepton number 
violating nature of the interaction underlying $\t1$ decays. The 
actual signal size may be considerably  larger due to reasons
discussed above.  Thus the  possibility that 
the reach will extend to higher $\MST$  or smaller $\epsilon_{br}$ 
is therefore quite open.

\begin{figure}[htb]
\centerline{
\includegraphics*[scale=0.6]{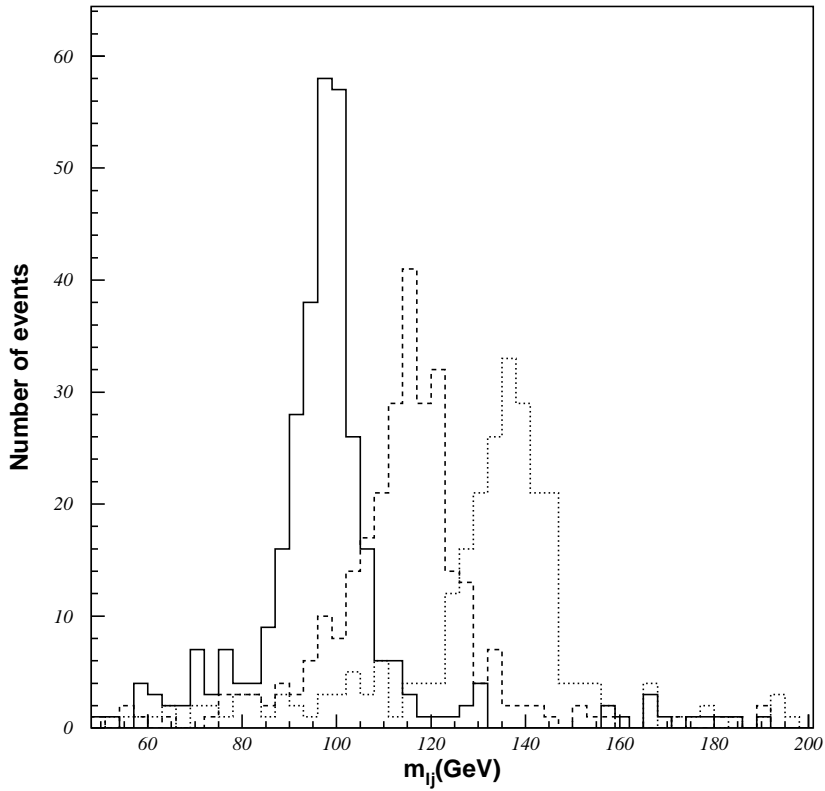}}
\caption{The lepton-jet invariant mass($m_{\ell j}$) distribution 
for ${\cal L}=$2~fb$^{-1}$,
for three top squark masses 100~GeV(solid line), 120~GeV(dashed line) 
and 140~GeV(dotted line).}
\label{lep_jj}
\end{figure}

\section*{4.~Stop decay branching ratios and the limits on $\lambda'$ in models
of $m_{\nu}$ }
\label{stopdecay}

As mentioned in the introduction
when the $\tilde t_1$ is the NLSP in  RPV  models of $m_\nu$,
three decay channels are allowed, which may naturally compete with
each other in various regions of the MSSM parameter space.
 They are the loop induced flavor changing decay mode,
Eq.(\ref{loopdk})~\cite{hikasa} the 4-body decay into 
states with nearly  massless fermions,
the bottom quark and the LSP,
Eq.(\ref{4bdk})~\cite{boehm,admgspd} and the RPV decay mode
Eq.(\ref{rpvmode}). In this section we discuss the  systematics of MSSM 
parameter space which enable us to identify the regions where 
different  decay modes dominate.

If the sleptons are lighter than  $\tilde t_1$, then the 3-body decay mode,
Eq.(\ref{3bdk}), involving sleptons open up. The competition between 
$\tilde t_1 \ar b \ell \tilde\nu$ and $\tilde t_1 \ar b \tilde\ell \nu $
and the RPV mode has been discussed in ref.~\cite{valle}. 
Here we shall also identify regions of the parameter space where the
decay modes given by Eq.(\ref{rpvmode}), (\ref{3bdk}) and 
(\ref{loopdk}) compete 
with each other. In the process we also demarcate the difficult regions
for Run-II in the context of top squark searches, where all decay modes 
may have relatively low rates.

It has been mentioned earlier that the couplings $\lambda'_{i33}$ are the most 
important ones in models of $m_\nu$. In ~\cite{abada} the upperbounds 
on these couplings were obtained from neutrino data in a variety of scenarii.
In all the cases the bounds were found to be approximately 
of the same order of magnitude ($\sim 10^{-3}-10^{-4}$). Our analysis 
based on the limiting BR of the last section can estimate the 
constraints on $\lambda'_{13j}$ or $\lambda'_{23j}$ that may 
arise from Run-II data. As before  we shall 
assume only one of these couplings to be dominating 
and shall henceforth drop the index of $\lambda'$.

All masses and mass parameters in this paper are in GeV.
For our analysis we fix the parameters: (i) The CP-odd neutral higgs 
mass $M_A$=300, which is required to calculate the $\epsilon$ parameter
(see Eq.(\ref{loopwidth})),  (ii) the trilinear coupling
in the sbottom sector $A_{b}$=300 and (iii) the trilinear coupling in the stau
sector $A_{\tau }$=200. The variation of the BR with respect to the 
 other parameters will be explicitly discussed 
as and when required.

\subsection*{4.1~Competition between the Loop induced 
and RPV decays}

As is well-known the loop decay width is controlled by the
parameter $\epsilon$ which denotes the amount of
$\tilde{t}_{L,R}$--$\tilde{c}_L$ mixing ~\cite{hikasa} and enters
in the decay width, 
\be
\Gamma(\tilde t_1 \ar c {\tilde\chi_1^0}) = \frac{\alpha}{4}
|\epsilon|^2 f^2 m_{\tilde t_1} \left(1 - \frac{m_{\tilde\chi_1^0}^2}{
m_{\tilde t_1}^2}\right)^2
\label{loopwidth}
\ee
where $f$ is the composition of neutralino mixing. 
The detailed expressions for $\epsilon$ and the function {\it f} can be found
in Ref.~\cite{hikasa,boehm}. Neglecting the lepton masses the decay width 
of the channel $\tilde t_1 \ar l^{+} d $ ( $\l$=e or $\mu$) is
\begin{equation}
\Gamma_{R\!\!\!\!/} = \frac{1}{16\pi} \lambda ^{\prime 2}
\MST cos\theta_{\tilde t}^2
\label{rpvdkwidth}
\end{equation}
where $\lambda'$ is the dominant RPV coupling and $\theta_{\tilde t}$ 
is the mixing angle in the top squark sector.

As long as the 2-body and 3-body RPC decay modes 
Eq.(\ref{2dk}) and Eq.(\ref{3bdk}) do not open up, i.e. if the top squark is 
the NLSP, the above two modes compete with each other.
In principle the 4-body decay mode could also enter into the competition.
However, in order to study the simplest  example of competing modes 
the latter has been suppressed by
considering relatively  large values of $\tb$. However, the competition 
among all these three decay modes will be considered later. 

If  $\lambda'$ is close to its current experimental bound from indirect
searches~\cite{gbherbi}, then the RPV decay dominates over the 
loop decay for the entire region of the parameter space unless
$\cost$ is fine tuned to be very small. The competition between the two
modes becomes  generic  when  $\lambda'$ is $\sim 10 ^{-3}- 10 ^{-4}$, 
which is interesting from the point of view of RPV models  of neutrino 
mass~\cite{numassnew}. The estimate $\lambda'$  $\sim 10 ^{-4}$, as 
mentioned in the introduction, is based on the assumption that the SUSY 
breaking scale ($ M_{SUSY}$) $\sim$ 100 ~\cite{abada}. Somewhat 
larger values of $ M_{SUSY}$ push this estimate
upwards. On the other hand value of  $\lambda'$ somewhat smaller than 
$\sim 10 ^{-4}$ may be relevant if the absolute values of the neutrino
masses, which are not known at the moment, are much smaller than the 
typical choice $\sim$ 1 eV.

As $M_2$ - SU(2) gaugino mass parameter(gaugino mass unification is 
assumed), $\mu$ - the higgsino mass parameter
and $\tb$ - the ratio of two vacuum expectation values of higgs sector,
completely describe the neutralino and the chargino sector. We have chosen 
these parameters such that the $\MCH$ is around 200, which more or 
less fixes the limit of the top squark mass up to which the competition 
among various decay channel can occur. Otherwise the 2 body decay mode of 
$\t1$, Eq.(~\ref{2dk}) will be open up and dominate
over all other decay modes. The common slepton mass is taken to be heavier 
than $\MST$ to avoid the 3-body decay channel.  

In Fig.~\ref{figa} the competition between these two decay mode
has been illustrated for various values of $\MST$. The other
MSSM parameters which are involved in this calculation are: $M_2$=250, 
$\mu$ = +250, $\tb$=40, the common scalar squark mass $m_{\tilde q}$=300, 
common slepton mass $m_{\tilde\ell}$=235, $\cost$=0.7 and $\lambda'$=0.001.

As  the top squark mass is increased, the $\eps$ parameter as well as the 
phase space factor
$ \left(1 - \frac{m_{\tilde\chi_1^0}^2}
{m_{\tilde t_1}^2}\right)^2 $ in Eq.(~\ref{loopwidth}) increase, 
but the former rises more sharply.
So, although both the widths in Eq.(~\ref{loopwidth}) and 
Eq.(~\ref{rpvdkwidth}) have a common
linear dependence on $\MST$, the loop decay BR dominates over
that of the RPV decay above a certain $\MST$.
This  happens for almost  all  
choices of the other  parameters, unless they are fine tuned to make
$\eps$ very small.

For smaller value of $\cost$, both the  loop and RPV decay widths
decrease, the former through the 
$\eps$ term and the latter through the direct dependence on $\cost$
respectively. The competition between the two BR still occur albeit
for higher top squark masses. The competition ceases to exist only if 
$\cost$ is fine tuned to make the $\eps$ parameter negligible.   

\begin{figure}[!htb]
\vspace*{-3.5cm}
\hspace*{-3.0cm}
\mbox{\psfig{file=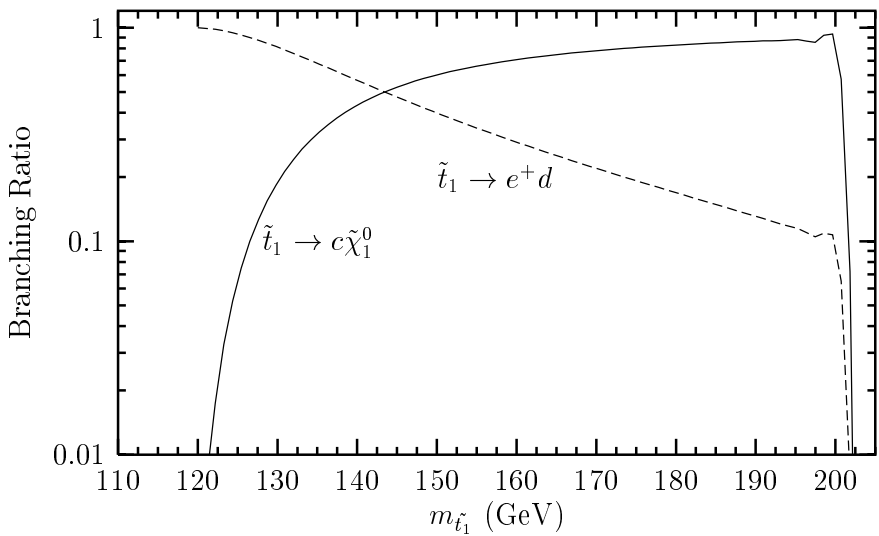,width=20cm}}
\vspace*{-16.7cm}
\caption{\small
The RPV and loop decay  BRs as functions of $\MST$. 
The Other MSSM parameters are $M_2$=250 , $\mu$ = +250 , $\tb$=40,
$m_{\tilde q}$=300 , $m_{\tilde\ell}$=235 , 
$\cost$=0.7 and  $\lambda'$=0.001.
}
\label{figa}
\end{figure}

\begin{figure}[!htb]
\vspace*{-3.5cm}
\hspace*{-3.0cm}
\mbox{\psfig{file=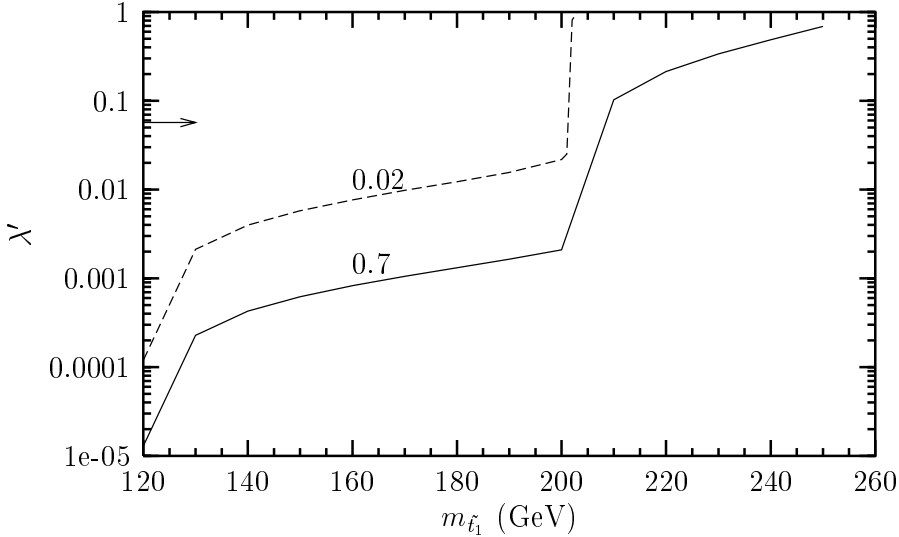,width=20cm}}
\vspace*{-16.7cm}
\caption{\small
The minimum value of $\lambda'$ at 5$\sigma$ for observable 
 signal for  $\cost$=0.7 (solid) and 0.02(dashed curve),
the others parameters are  the same as in  Fig.~\ref{figa}
}
\label{figa_excl}
\end{figure}

The RPV decay width depends on the product of $\cost$ and $\lambda'$.
Keeping this product ( i.e. the width in Eq.(\ref{rpvdkwidth})) fixed,
if we increase $\lambda'$, the loop width will decrease  
as a consequence of lowering 
$\cost$. So, the competition will take place for higher top squark  
masses only. However, above a certain  $\lambda'$ the loop decay fails to 
compete for the entire range of $\MST$ corresponding to a
top squark NLSP. On the other hand for smaller $\lambda'$ the RPV decay 
width is scaled down in a straight forward way. Now the competition 
occurs over a larger range of $\MST$ and for smaller values of $\cost$ and 
/or $\tb$.

If $\tb$ is lowered, for fixed $\mu$ and $M_2$, the chargino mass
is lowered by a small amount so that the threshold for the 2-body
decay is slightly lowered. More importantly, the $\eps$ parameter
decreases dramatically below a certain $\tb$. Here the RPV decay
overwhelms the loop decay. However, precisely for such low values of 
$\tb$ the 4-body decay become important if, in addition, the chargino is
of low virtuality ($\MST$ $\approx$ $\MCH$; see the next subsection)

Even if $\lambda'$ is as low as $\sim 10^{-4}$ the competition between 
the two modes still exists for smaller values of $\tb$ which lowers 
$\eps$ and, hence, the loop decay width. 
The minimum value BR, as shown in Fig.\ref{fig_stop} 
can be traded to find the 
limiting value of $\lambda'$ considering Eq.(~\ref{loopwidth}) and 
\ref{rpvdkwidth}.  In Fig.~\ref{figa_excl} 
the two curves represent limiting values of $\lambda'$ for observable 
signal for two values of 
$\cost$. The regions above the curves corresponds to observable 
BR as given in Fig.\ref{fig_stop}.
The other SUSY parameters chosen are as in Fig.~\ref{figa}.
In this figure and the similar ones presented subsequently  
the horizontal arrow represents the upper bound on $\lambda'_{131}$
prior to the neutrino data~\cite{gbherbi}. The bounds 
on $\lambda'_{132}$ and $\lambda'_{23j}$ are even weaker. Only the 
bound on $\lambda'_{133}$ is $\sim 10^{-3}$. Hence significant 
improvement in the existing limits on many RPV couplings is expected. 
For larger $\cost$, the RPV width increase significantly. As a result 
the BR constraint is satisfied for lower $\lambda'$. The sharp rise in 
the curve for $\MST$ $\gsim$200 is a consequence of the opening up of the 
2-body channel $\tilde t_1 \ar b \ch1pm $. It is interesting to note that 
for large $\cost$ the data will be sensitive to the values of $\lambda'$ 
relevant for neutrino masses until the 2-body decay channel opens up.

\subsection*{4.2~Competition between the 4-body and RPV decay}

The dependence of the 4-body decay rate on supersymmetric parameters
has been discussed in great detail in  
Ref.~\cite{boehm,admgspd}. The competition between 4-body decay modes
with the loop induced flavor changing decay mode Eq.(\ref{loopdk})
has been discussed both in MSSM and mSUGRA models in 
Ref.~\cite{admgspd}.

In general in order to identify the parameter space relevant for
the competition between the RPV decay channel 
and the 4-body decay channel, one has to
take the $\t1$ to be almost right handed (i.e., $\cost$ small)
and $\lambda' \sim 10^{-3}$ or $10^{-4} $. For small value of $\tb$, 
the loop decay amplitude becomes negligible without
requiring any fine tuning. 
In Fig.~\ref{figb} we demonstrate this competition
for  $\lambda' = 10^{-4}$
as a function of $\MST$. The choice of the MSSM parameters are 
explicitly mentioned in the figure caption.

\begin{figure}[!t]
\vspace*{-3.5cm}
\hspace*{-3.0cm}
\mbox{\psfig{file=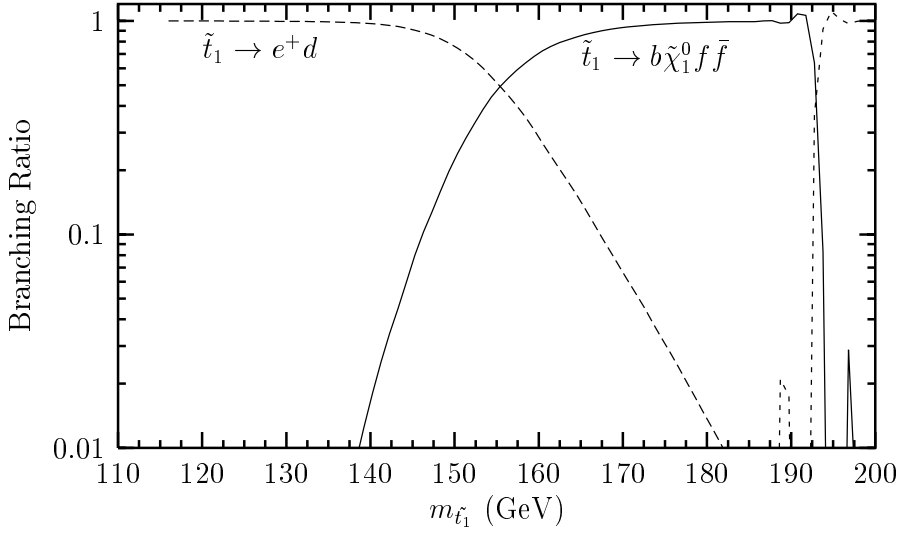,width=20cm}}
\vspace*{-16.7cm}
\caption{\small
The RPV and 4-body decay BRs are shown. 
The Other MSSM parameters are $M_2$=250 , $\mu$ = +250 , $\tb$=6,
$m_{\tilde q}$=300 , $m_{\tilde\ell}$=210 , 
$\cost$=0.1 and  $\lambda'$=0.0001.
}
\label{figb}
\end{figure}

\begin{figure}[!t]
\vspace*{-3.5cm}
\hspace*{-3.0cm}
\mbox{\psfig{file=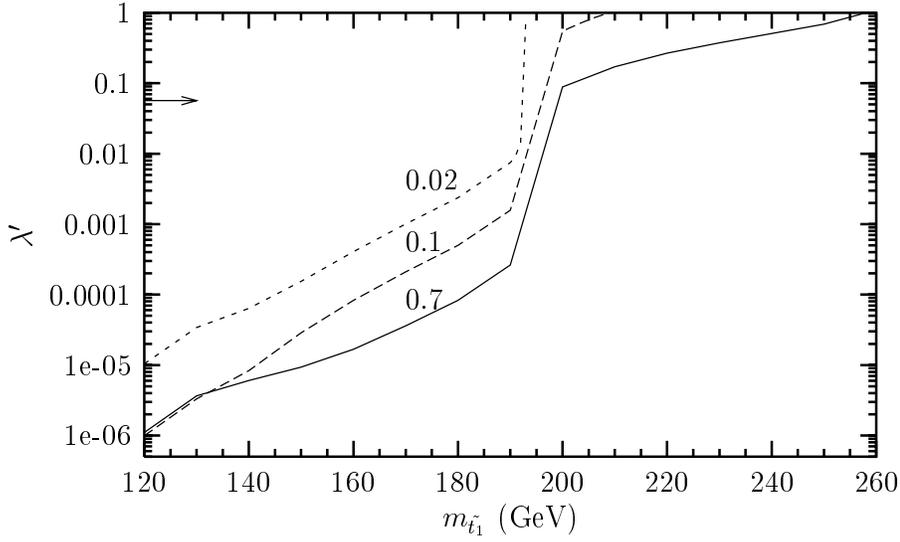,width=20cm}}
\vspace*{-16.7cm}
\caption{\small
The minimum value of $\lambda'$ at 5$\sigma$
for observable signal for 
$\cost$=0.7(solid), 0.1(dashed) and 0.02 ( dotted curve). The other 
parameters are the same as in Fig.~\ref{figb}}
\label{figb_excl}
\end{figure}

As expected the 4-body decay channel opens up for relatively low
($\MST$ - $\MCH$ )
so that the chargino in the 4-body decay process has a small virtuality.
As before, in Fig.~\ref{figb_excl} we show the potential region(upper side) 
of $\lambda'$
which can be probed at Run-II experiments. This region 
corresponds to the BR constraint(Fig.~\ref{fig_stop}) are shown for  
three values of $\cost$ and other SUSY parameters chosen are as in
Fig.~\ref{figb}. The conventions are same as in Fig.~\ref{figa_excl}.

\subsection*{4.3~Competition between the Loop, 4-body and RPV decay }

In order to illustrate  the possibility of  
competition between the above  three channels, we shall keep in mind 
that the $\epsilon$ parameter 
must not be as small as in the previous section. 
The competition is demonstrated  in Fig.\ref{figc} with the 
choice of SUSY parameters mentioned explicitly in the figure
caption.

\begin{figure}[!t]
\vspace*{-3.5cm}
\hspace*{-3.0cm}
\mbox{\psfig{file=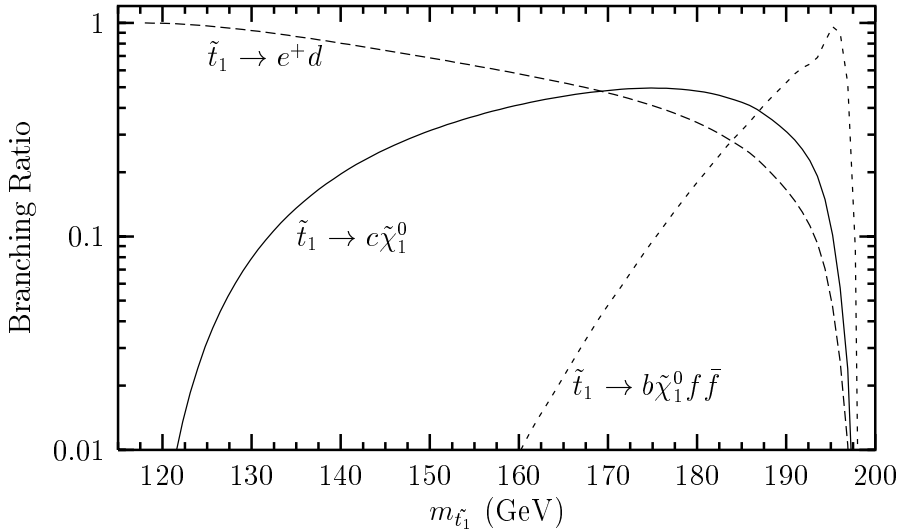,width=20cm}}
\vspace*{-16.7cm}
\caption{\small
The RPV, loop and 4-body decay BRs are shown. 
The other MSSM parameters are $M_2$=250 , $\mu$ = +250 , $\tb$=10,
$m_{\tilde q}$=300 , $m_{\tilde\ell}$=210 , 
$\cost$=0.9 and  $\lambda'$=0.0001.
}
\label{figc}
\end{figure}

\begin{figure}[!t]
\vspace*{-3.5cm}
\hspace*{-3.0cm}
\mbox{\psfig{file=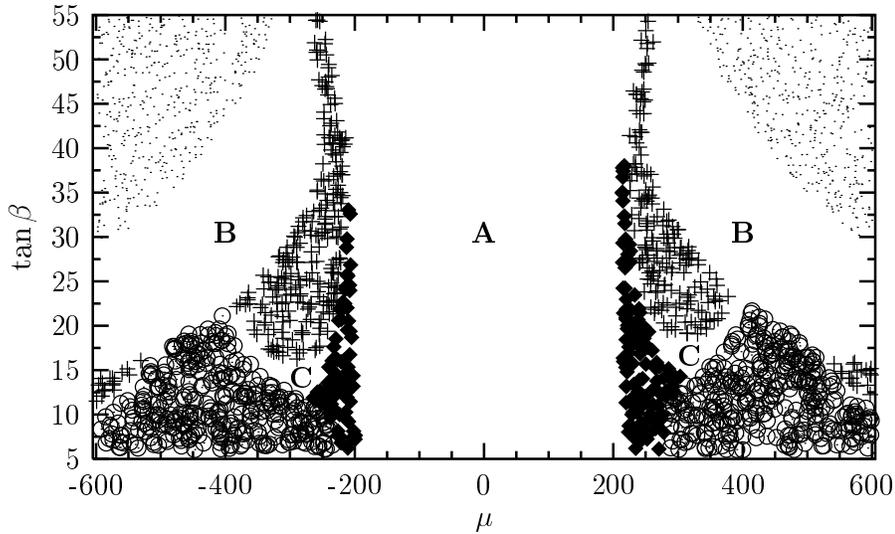,width=20cm}}
\vspace*{-16.7cm}
\caption{\small
Different regions dominated by a particular decay channel of top squark 
are shown for $\MST$=180~. See the text for conventions for   
demarcating regions. Except for $\cost$=0.3, all the other 
parameters are the same as in
Fig.~\ref{figc}. 
}
\label{fig6_mutanb}
\end{figure}

\begin{figure}[!t]
\vspace*{-3.5cm}
\hspace*{-3.0cm}
\mbox{\psfig{file=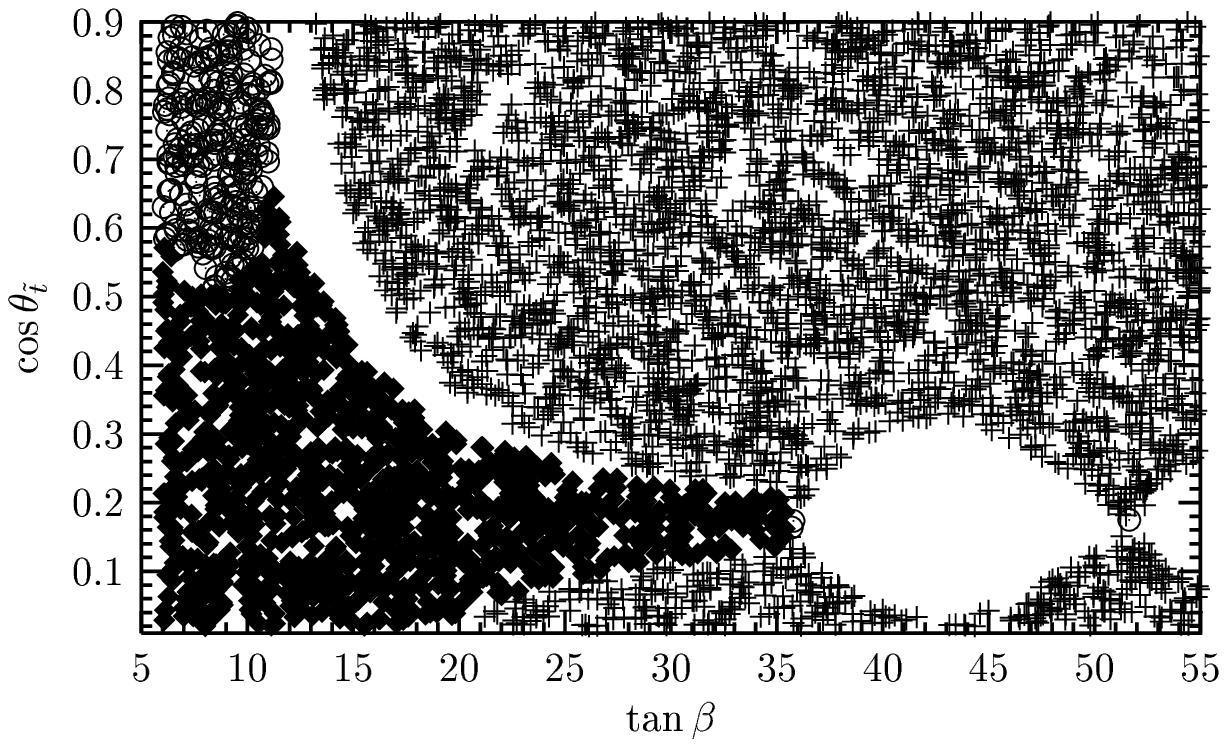,width=20cm}}
\vspace*{-16.7cm}
\caption{\small
Similar information as is in Fig.~\ref{fig6_mutanb} using different variables. 
}
\label{fig6_costtanb}
\end{figure}

\begin{figure}[!t]
\vspace*{-3.5cm}
\hspace*{-3.0cm}
\mbox{\psfig{file=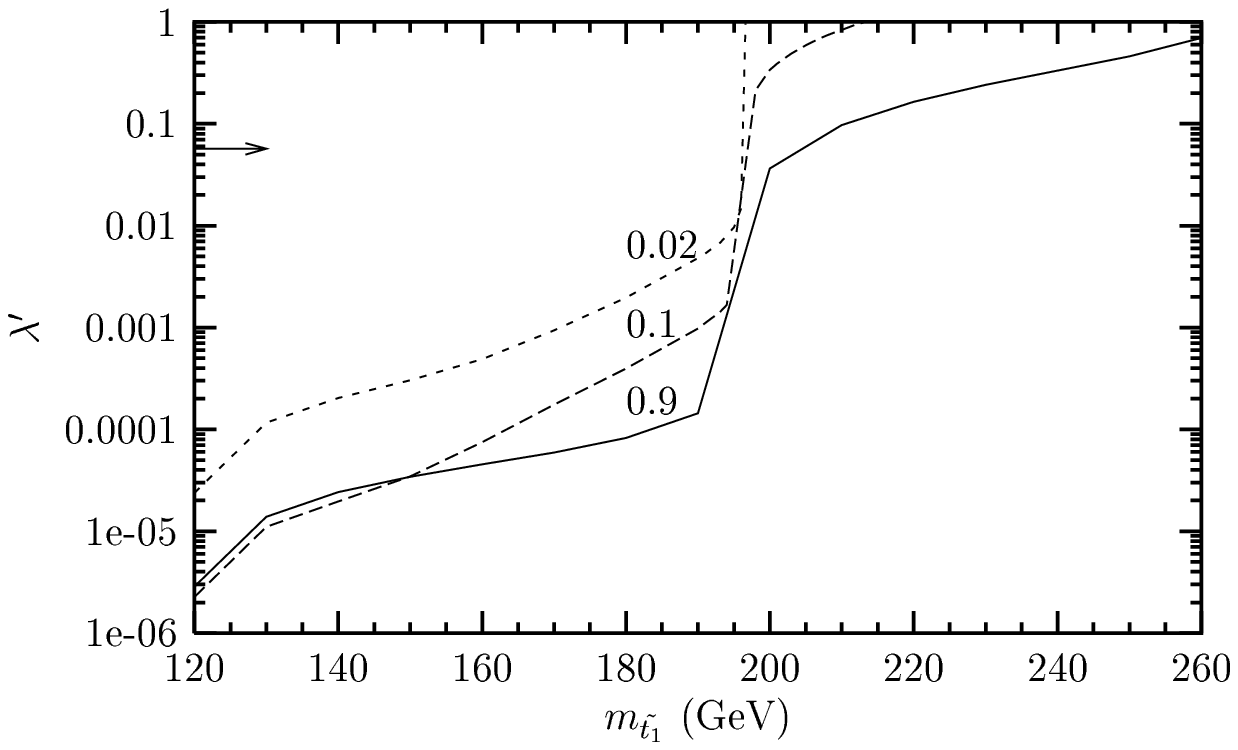,width=20cm}}
\vspace*{-16.7cm}
\caption{\small
Similar information as in Fig.\ref{figb_excl},  for  
$\cost$=0.9(solid), 0.1(dashed) and 0.02(dotted curve).
 Other parameters are the same as in  Fig.\ref{figc}
}
\label{figc_excl}
\end{figure}
The total 4-body BRs is significant ($\gsim$ 10\%) for the range 
of top squark mass,
$\MCH$ - 20 $\lsim$ $\MST$ $\lsim$ $\MCH$. In this top squark mass range 
the chargino is of very small virtuality. Also
we choose $m_{\tilde\ell}$ i.e. the common slepton mass, such that even after 
the mixing in the third generation, the lighter tau slepton mass
$\MSTAU$ is above  the chargino mass. So the slepton mediated 4-body process
also  has a low virtuality. But 
the 3-body decay mode Eq.(~\ref{3bdk}) is still kinematically forbidden.

If the signal is seen in all three channels then one has to 
identify the region of the parameter space where the corresponding 
BRs are above the observable limit. Similarly
in order to exclude a particular $\MST$ comprehensively it is essential
to establish that at least one of the competing modes would be observable 
over the entire parameter space. In order to do a complete job one needs
the minimum observable BRs at Run-II for each of the allowed modes.
Unfortunately at the moment we have numerical 
estimates for the RPV mode ($\tilde t_1 \ar e^+ d_j$) only
( Fig.\ref{fig_stop}). In the following we shall delineate the regions
of the parameter space where i) the RPV decay is observable at Run-II
or ii) either one of the two competing modes have  sizable BR.

The relevant information will be presented in the form of scatter plots
obtained by varying two important parameters randomly keeping
the others fixed. The scatter plots
also illustrate the competition among the decay modes in 
specific regions of the parameter space.

In Fig.\ref{fig6_mutanb}, fixing $\MST$=180, $\cost$=0.3, $\mu $ and $\tb$
are varied randomly setting the other parameters as in Fig.\ref{figc}. 
The fixing of $\MST$, which tacitly assumes
that $\MST$ can be reconstructed, make the analysis simpler.
The systematic of the parameter space is clear from
Fig. \ref{fig6_mutanb}.
In this figure, the regions marked by the circle is the one where 
the RPV mode is above the observable limit, i.e.,  
BR( $\tilde t_1 \ar e^+ + d$) 
$\gsim$ 26\%(see Fig.1). Although the width of this mode does not 
depend upon $\mu$ or $\tb$ directly, however, its BR is quite 
sensitive to these parameters. The regions marked by '{\bf + }'
correspond to BR( $\tilde t_1 \ar c \LSP $)$\gsim$ 75\% 
and that marked 
by the black diamonds correspond to 
BR ($\tilde t_1 \ar b \tilde\chi_1^0 f \bar f'$) $\gsim$ 30\%
with the RPV BR less than the observable limit. Note that for low values of 
$\tb$, the RPV BR is much larger than 30\%. The region 
labeled by '{\bf A}', $\MST > \MCH + m_b$, the 2-body decay 
mode Eq.(\ref{2dk}) opens up and overwhelms all other 
decay channels. Finally, the region marked by '{\bf B}',
where $\mu$ and $\tb$ are large, the lighter $\tau$ slepton mass 
eigenstate ($\STAU$) becomes rather light and the 3-body  decay mode
involving a $\STAU$ in the final state strongly dominates.
In the dotted region $\STAU$ is lighter than the $\LSP$ or has 
unphysical mass. Although $\MSTAU \lsim \MLSP$ is allowed in RPV MSSM 
in general, but we are not interested in top squark signals in this
scenario. 

If the top squark signal is seen in one 
or more channels then one can broadly identify 
the relevant region of the parameter space. For example
if all the three modes are seen then the white region 
('{\bf C}')  or regions in its neighborhood may  be of interest.
For more precise conclusion one needs to know the 
limiting BR of all the modes quantitatively . One can hope that Tevatron 
Run-II and/or LHC will gradually supply the relevant information. 
However, the same region  may eventually turn out to   be 
the difficult one  to exclude at Run-II, if no signal is seen, 
since in parts of  this region all the BRs may turn out to
be below the observable limit.

In  Fig. \ref{fig6_costtanb} the scatter plot is in $\tb -\cost$ plane
with $\mu$=250 and the other parameters as in Fig.\ref{figc}. The convention
for demarcating the regions are the same as in  Fig.\ref{fig6_mutanb}. 
Again the white regions could be the difficult ones from the point
of view of comprehensive $\t1$ search at Run-II.

In Fig.~\ref{figc_excl} the three curves represent limiting 
values of $\lambda'$  for observable signal for three values of 
$\cost$. 
The regions above the curves corresponds to observable 
BR as given in Fig.\ref{fig_stop}.
The other SUSY parameters chosen are as in Fig.~\ref{figc}.
It is interesting to note that  for large $\cost$ the data will 
be sensitive to values of  $\lambda'$ relevant for neutrino masses 
until the 2-body decay channel opens up.
It is to be noted that  for relatively small $\MST$ the bound is 
fairly insensitive to $\cost$ for the range 0.1$\lsim $ $\cost$$ \lsim$ 0.9.
As the  threshold of the 4-body decay open up for larger $\MST$, 
the constrain on  $\lambda'$ gets weaker as expected.

\subsection*{4.4~Competition between the 3-body, 
Loop and RPV decay }
The competition between RPV decay mode, the loop decay,  and all 
RPC 3-body channels 
has been studied in ~\cite{3bdkexpli}. In this section
 we consider a scenario where top squark is not the NLSP and the 
first two RPC decay modes of Eq.(~\ref{3bdk}) are open.
 We have then studied the competition
among these two  modes, the loop decay and the RPV decay  
taking into account the limiting BR of the last mode
 obtained in Sec. 3.
 As the 3-body decay mode is  kinematically allowed for light 
sleptons only, the slepton mass should be  chosen with care so that it is
consistent with the 
experimental lower limit.

\begin{figure}[!t]
\vspace*{-3.5cm}
\hspace*{-3.0cm}
\mbox{\psfig{file=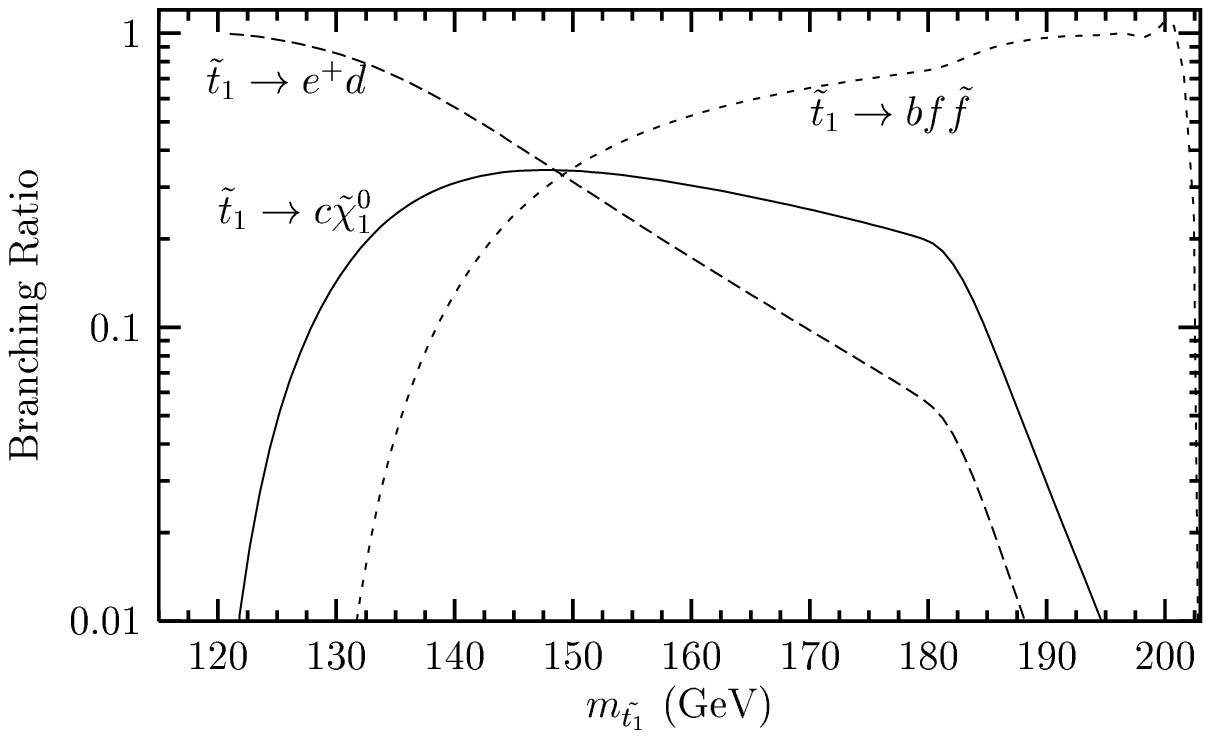,width=20cm}}
\vspace*{-16.7cm}
\caption{\small
The RPV, loop  and 3-body decay  BRs as functions of $\MST$. 
The other MSSM parameters are $M_2$=250 , $\mu$ = +250 , $\tb$=40,
$m_{\tilde q}$=300 , $m_{\tilde\ell}$=175 , 
$\cost$=0.5 and  $\lambda'$=0.001.
}
\label{figd}
\end{figure}

\begin{figure}[!t]
\vspace*{-3.5cm}
\hspace*{-3.0cm}
\mbox{\psfig{file=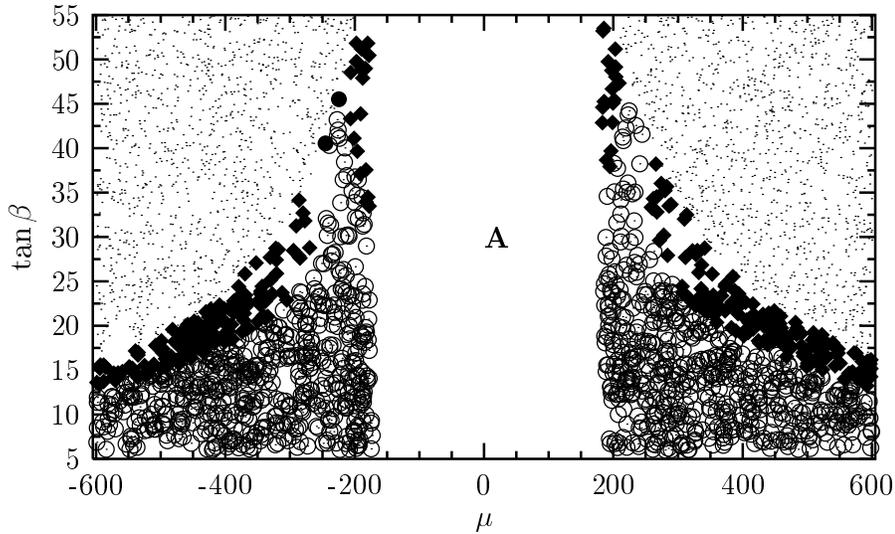,width=20cm}}
\vspace*{-16.7cm}
\caption{\small
Similar information as in Fig.\ref{fig6_mutanb}, but for $\MST$=160 .
The other parameters are the same as in
Fig.~\ref{figd}.}
\label{fig8_mutanb}
\end{figure}
\begin{figure}[!t]
\vspace*{-3.5cm}
\hspace*{-3.0cm}
\mbox{\psfig{file=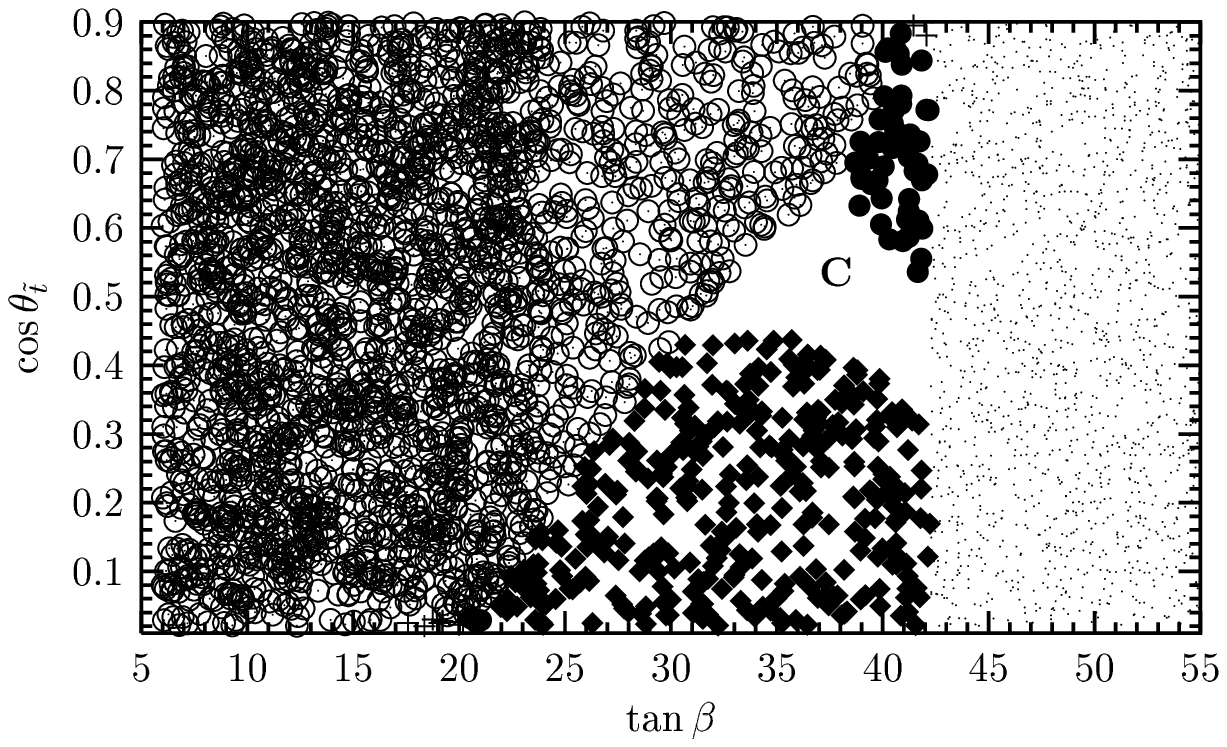,width=20cm}}
\vspace*{-16.7cm}
\caption{\small
Similar information  as in Fig.\ref{fig8_mutanb} using different variables.
}
\label{fig8_costtanb}
\end{figure}

\begin{figure}[!t]
\vspace*{-3.5cm}
\hspace*{-3.0cm}
\mbox{\psfig{file=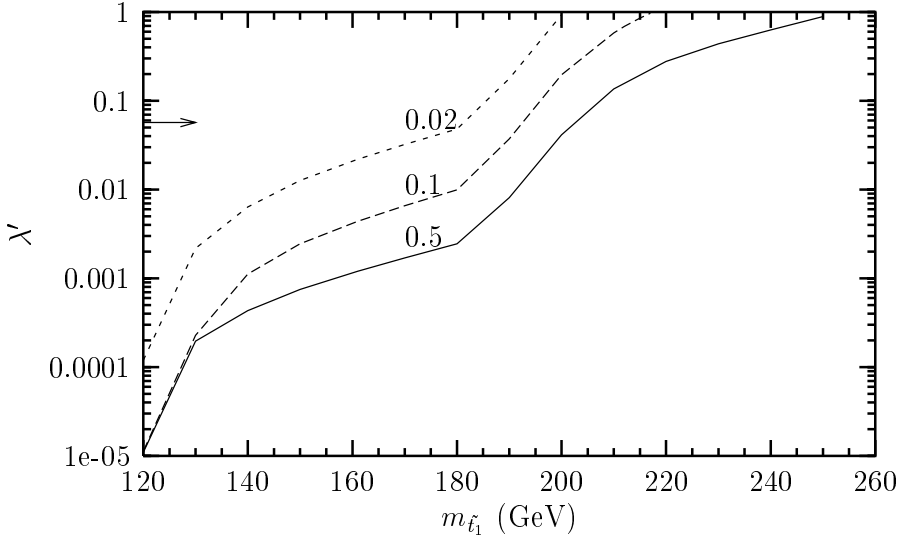,width=20cm}}
\vspace*{-16.7cm}
\caption{\small
Similar information as in Fig.\ref{figb_excl} for  
$\cost$=0.5(solid), 0.1(dashed) and 0.02(dotted curve)
and the others parameters are the  same as in  Fig.~\ref{figd}
}
\label{figd_excl}
\end{figure}

With the choice of the SUSY parameters as in 
Fig.\ref{figd}, the  3-body decays, if kinematically 
allowed, have BR $\gsim$ 10\%  almost for the  entire range 
of top squark masses. For this set of parameters, the $\MSTAU$,
$\MSNU$, and $\MSL1$ ($l=e$ or $\mu$)  are 124 , 156  and 
175 respectively.
 
Interestingly we have found that when the chargino is in the mixed
region with a relatively large mass, i.e.,  when the 3-body decay width
is somewhat reduced both due to a mixing angle factor and propagator
suppression, there may be a competition among the loop, 3-body decay 
and RPV decay modes for 
$\lambda'$=0.001. Here the loop decay width is significant
thanks to relatively large $\cost$=0.5 and large $\tb$. This is demonstrated
in  Fig.\ref{figd}. It follows from Fig.\ref{figd} that in the
neighborhood
of $\MST$=150 all three modes coexist with nearly equal BR. However
from the
limiting  BR  plot( see Fig.\ref{fig_stop} ) we find that in this region 
the signal is observable if BR($ \tilde t_1 \ar e^+ + d $ ) $\gsim$20\%. 
Hence the loop decay channel may not be very important as the discovery 
channel. If the BR($ \tilde t_1 \ar e^+ + d $ ) is below the 
observable limit, the 3-body mode will be the main discovery channel.
In Fig.\ref{figd} for relatively low $\MST$ the 3-body mode with
$\STAU$ in the final state opens up. For the higher $\MST$ the modes
with $\SNU$ and other sleptons in the final state are allowed.

In Fig.\ref{fig8_mutanb} the competition among the decay modes is
illustrated in $\mu-\tb$ plane, for $\MST=$160~. For this $\MST$
only the 3-body mode with the $\STAU$ in the final state is relevant.
Depending on the values of $\tb$, $\mu$ and their product in different 
regions of the parameter space the competing decay modes are kinematically 
allowed. In Fig.\ref{fig8_mutanb} the dotted 
circles delineate the parameter space where the RPV decay is observable.
The regions characterized by relatively large $\mu$ and $\tb$ corresponds to 
the light $\STAU$ scenario. This part of the parameter space dominated by
the 3-body decays ( BR $\gsim$ 70\% 0 is marked
with the black diamonds. The dotted region
is theoretically disfavored as explained in the context of Fig.8.  
Finally, the  black circles represent the parameter space with 
40\% $\lsim$ BR($\tilde t_1 \ar c \tilde\chi_1^0$)$\lsim$70\%. 
Only a few points appear at large $\tb$. In the region marked with '{\bf A}'
the 2-body decay overwhelms the other modes. In  Fig.\ref{fig8_costtanb}
a $\tb-\cost$ scatter plot is presented following the same
convention.

In Fig.\ref{figd_excl} we present the limiting value of $\lambda'$
for three values of $\cost$ corresponding to the parameter space
of Fig. \ref{figd}.  The first change in the slope occurs
at about $\MST$=130 due to the opening up of the channel
$\tilde t_1 \ar b \nu_\tau \STAU$. The second change in the neighborhood of
$\MST$=180 corresponds to the decay mode
$\tilde t_1 \ar b \nu_\l \tilde \l_1$ ( $l=e$ or $\mu$) .
In both cases the BR of the  RPV decay mode are reduced which have to be
compensated by higher values of $\lambda'$.
Finally for $\MST \gsim $ 210, the
2-body channel, Eq.(\ref{2dk}), becomes the main decay mode.

\section*{5.~Conclusion }

It is quite possible that the mass of  the lighter top squark is much
smaller than the 
other squarks and gluinos due to mixing and RG effects and it is the only 
strongly 
interacting superparticle within the kinematic reach of Run-II of the 
Tevatron with large production cross-section. If this is the case then the 
RPV decay  $\tilde t_1 \ar l_i^+ d_j$ driven by the trilinear coupling 
$\lambda'_{i3j}$, where $i$ and $j$ are generation indices, may be the most 
attractive channel for discovering R-parity 
violation\cite{asesh,valle,subhendu}.

Additional interest in this process stems from the fact that some subset
of the above couplings, in particular $\lambda'_{i33}$,  may 
be important ingredients of RPV
models of $m_\nu$\cite{numassnew,abada}.
This scenario constraints the magnitudes of these couplings
to be generically small ($\lsim 10^{-3} - 10^{-4} $, see, 
e.g., \cite{abada}).

If the couplings are indeed so small the RPC 2-body decay, (Eq.(\ref{2dk}))
or 3-body decay modes (Eq.(\ref{3bdk})), if kinematically allowed, would overwhelm 
the RPV decay and the LSP decay may be the only 
signature of R-parity violation \cite{barger}. This signature,
Eq.(\ref{lsp3bdk}), however, may not reveal the lepton number violating
nature of the underlying interaction or whether the strength 
of the coupling is indeed in the right ballpark required by models 
of $m_\nu$.

The situation is dramatically different if the $\t1$ is the NLSP since
the allowed RPC decays - the loop induced ( Eq.(\ref{loopdk}))
or the 4-body ( Eq.(\ref{4bdk})) channel - are naturally suppressed.
If the RPV coupling is indeed $ \sim 10^{-3} - 10^{-4} $ then 
BR of the three allowed channels may indeed be comparable. Thus
the simultaneous observation of two or more of these decay may be 
a hallmark of RPV models of $m_\nu$.

In Sec.3 using event generator {\tt PYTHIA}\cite{spythia} we have 
estimated the minimum value of BR of the RPV decay channel 
$\tilde t_1 \ar e^+ d_j$ for various values of $\MST$ 
corresponding to observable signals at Run-II experiments. 
Our results (see Fig.\ref{fig_stop}) show that much smaller 
BR can be probed at Run-II with 2$fb^{-1}$ of data compared 
to the bounds obtained from Run-I data \cite{subhendu}. These
results are approximately valid for down type quarks of
all generations and also for the channel 
$\tilde t_1 \ar \mu^+ d_j$. In reality the limiting 
BR may be much smaller than our conservative estimates as can be seen
by using enhanced NLO cross-section\cite{spira}, larger integrated luminosity 
or by employing b-tagging, since  in many models $\lambda'_{i33}$  are the
most important couplings, to improve the S/$\sqrt B$ ratio. Our simulations
show that $\MST$ can be reconstructed from the decay products with
reasonable accuracy, revealing thereby the lepton number violating nature of
the underlying decay dynamics.  

It is gratifying to note that even our conservative 
estimates of the limiting BR can be translated into interesting 
upper bounds on the RPV couplings $\lambda'_{i3j}$ ($i$=e or $\mu$) 
for representative choices of the MSSM parameters if no signal is 
seen (Figs. \ref{figa_excl}, \ref{figb_excl},
\ref{figc_excl},\ref{figd_excl}).  Thus the existing bounds \cite{gbherbi}
on several $\lambda'_{i3j}$ ( except perhaps $\lambda'_{133}$ ) can be
significantly improved. 
These results indicate that the Run-II data will indeed be sensitive to
magnitudes of these couplings even if they are  as small as that required 
by the models of $m_\nu$.

Using our estimate of the limiting BR as a function of $\MST$
one can demarcate the regions of the MSSM parameter space 
in specific models, where the 
RPV decay is observable. In Sec. 4 we have also studied the systematics of the 
MSSM parameter space and have 
delineated the regions where the competing decay modes are numerically 
significant. One 
can also have some idea of the difficult regions  of the parameter
space where the BR of none of the competing decays clearly dominates. 
All these information will become more precise 
once full simulations of the competing signals 
(Eq.(\ref{loop_lspdk}) and Eq.(\ref{4bd_lspdk})) 
estimate the limiting BR corresponding to all signals .
If no signal is seen then programme for top squark search at the LHC
may focus on the regions of the parameter space, which were difficult
in Run-II experiments.

\noindent
{\bf Acknowledgements :} AD acknowledges financial support from 
BRNS(INDIA) under the project number 2000/37/10/BRNS.
SPD acknowledges the grant of a senior fellowship by CSIR, India. He also 
thanks N.K.Mondal and members of the DHEP group for providing 
support during the visit at Tata Institute of Fundamental Research 
where part of this work was done.

\end{document}